\ifdefined\ONECOLUMN
\documentclass[journal, 12pt, onecolumn, draftclsnofoot, letterpaper]{IEEEtran}
\else
\documentclass[journal, a4paper, oneside]{IEEEtran}
\fi

\IEEEoverridecommandlockouts

\usepackage{amsmath,amssymb,mathrsfs}
\usepackage{cite}
\usepackage{graphicx,color}
\usepackage{hyperref}
\usepackage{psfrag}
\usepackage{subfigure}
\usepackage{url}
\usepackage{balance}
\usepackage{stfloats}

\usepackage{array}
\usepackage{fancyhdr}
\usepackage{float}
\usepackage{epsfig}
\usepackage[nomain,acronym,toc]{glossaries}
\usepackage{algorithm,algorithmic}
\usepackage{caption}
\newtheorem{thm}{Theorem}
\newtheorem{cor}{Corollary}[thm]

\newtheorem{lem}{Lemma}

\renewcommand{\eqref}[1]{(\ref{#1})}
\definecolor{sblue}{RGB}{0,51,160}

\ifCLASSINFOpdf
\else
\fi

\begin{document}
\title{Massive-MIMO MF Beamforming with or without  Grouped STBC for Ultra-Reliable Single-Shot Transmission Using Aged CSIT}
\author{Jinfei Wang, Yi Ma, Na Yi, Rahim Tafazolli, and Zhibo Pang
\thanks{Jinfei Wang, Yi Ma, Na Yi, and Rahim Tafazolli are with the 5GIC and 6GIC, Institute for Communication Systems, University of Surrey, Guildford, United Kingdom, GU2 7XH, e-mail: (jinfei.wang, y.ma, n.yi, r.tafazolli)@surrey.ac.uk. }
\thanks{Zhibo Pang is with the Department of Automation Technology, ABB Corporate Research Sweden, Vasteras, Sweden, and the Department of Intelligent Systems, Royal Institute of Technology (KTH), Stockholm, Sweden, email: (zhibo@kth.se). ({\it Corresponding author: Yi Ma})}
}


\maketitle

\begin{abstract}	
The technology of using massive transmit-antennas to enable ultra-reliable single-shot transmission (URSST) is challenged by the transmitter-side channel knowledge (i.e., CSIT) imperfection. When the imperfectness mainly comes from the channel time-variation, the outage probability of the matched-filter (MF) transmitter beamforming is investigated based on the first-order Markov model of the aged CSIT. With a fixed transmit-power, the transmitter-side uncertainty of the instantaneous signal-to-noise ratio (iSNR) is mathematically characterized. In order to guarantee the outage probability for every single shot, a transmit-power adaptation approach is proposed to satisfy a pessimistic iSNR requirement, which is predicted using the Chernoff lower bound of the beamforming gain. Our numerical results demonstrate a remarkable transmit-power efficiency when comparing with power control approaches using other lower bounds. In addition, a combinatorial approach of the MF beamforming and grouped space-time block code (G-STBC) is proposed to further mitigate the detrimental impact of the CSIT uncertainty. It is shown, through both theoretical analysis and computer simulations, that the combinatorial approach can further improve the transmit-power efficiency with a good tradeoff between the outage probability and the latency.
\end{abstract}

\begin{IEEEkeywords}
	Ultra-reliable single-shot transmission (URSST), massive multiple-input multiple-output (massive-MIMO), spatial diversity, matched-filter (MF) beamforming, grouped space-time block codes (G-STBC).
\end{IEEEkeywords}

\IEEEpeerreviewmaketitle


\section{Introduction}\label{secI}
Mission-critical wireless applications (e.g., industrial private networks, public safety networks, and vehicle-to-everything (V2X) networks) require ultra-reliable low-latency communications (URLLC) \cite{8663456,8341501,9088229,8630650}. 
Some of them are even demanding for a short data packet to be conveyed for only once while ensuring an ultra-low packet-error-rate (PER) \cite{8660457,8723572}. 
This is termed the ultra-reliable single-shot transmission (URSST), which has recently been considered by Nokia Bell Labs as one of key technologies towards 6G (see \cite{9114878}). 
The stringent requirements of URSST are driving the extreme use of available physical channel degrees-of-freedom (DoF) for the diversity gain.
Moreover, the transmitter must form a reasonable prediction of the instantaneous signal-to-noise ratio (iSNR) and conduct power adaptation accordingly. 
For practical cases when an accurate iSNR prediction is not possible, the transmitter can form a pessimistic prediction instead as long as it can guarantee the target PER without compromising too much of the transmit-power efficiency \cite{8329618}. 

Massive multiple-input multiple-output (MIMO) is a scalable multi-antenna transmission technology for the exploitation of the channel spatial DoF \cite{6375940}. 
Given the perfect knowledge of channel-state information at the transmitter (CSIT), a low-complexity matched-filter beamforming can maximize the iSNR for a single-antenna receiver \cite{Tse2005}. 
When the number of transmit-antennas tends to infinity, the iSNR fluctuation tends to zero in spite of fading channels. 
This is the well-known channel hardening effect inherent in massive-MIMO, which makes the iSNR almost certain to the transmitter \cite{Hochwald2004}.
All of these are encouraging the use of massive-MIMO technology for URSST particularly when a wireless system has very limited temporal and/or frequency DoFs.
Despite, massive-MIMO is challenged by the CSIT imperfection, which can be due to various reasons such as channel estimation inaccuracy, channel feedback delays, or not up-to-date channel knowledge, etc. 
The CSIT imperfectness can be less detrimental to eMBB services as they are generally optimized for the average throughput or the average SNR (aSNR).  
However, it is a critical problem for the URSST since the CSIT imperfectness can result in unreliable iSNR prediction, which is very detrimental to the task of delivering ultra-reliability services such as $10^{-5}$ or lower PER for every single packet transmission. 
In addition, throughput-oriented massive-MIMO systems are not suitable for the URSST as the latter is extremely demanding for the reliability. 
Therefore, the current massive-MIMO technology must be fundamentally re-devised in order to fulfill the URSST requirements. 

In this work, we are interested in a wireless system where the transmitter employs massive antennas to deliver URSST services to mobile terminals.
To ensure the ultra-reliability and low latency, it is practical to assume users sharing their wireless medium orthogonally in frequencies. 
Such could ensure no inter-user interferences in the spatial domain and allow the transmitter to exploit all of available spatial DoF for the diversity gain. 
It is also assumed that the transmitter has some level of channel knowledge, which can be obtained either through the channel reciprocity in time-division duplexing (TDD) or channel feedback in frequency-division duplexing (FDD). 
Due to channel time-variations, the channel knowledge is considered to be the aged CSIT, which is described by the first-order Markov model.
For an interference-free system, the low-complexity matched-filter (MF) beamforming is deemed to be the SNR-optimized approach, which is therefore of our main interest in this work. 

The idea of utilizing massive-MIMO beamforming for the URLLC is not novel. 
There are already some recent contributions in place, which aim to exploit the massive-MIMO channel hardening effect to mitigate the negative impact of fading channels (e.g., \cite{9399808,8707069,9107489,8673808,9416241}). 
For the closed-loop MIMO, the quality of CSIT is important to the massive-MIMO beamforming, particularly for URLLC applications. 
The impact of channel estimation inaccuracy on massive-MIMO-enabled URLLC as well as their corresponding mitigating strategies have been partially studied in the literature (e.g., \cite{8640115,9044874}). 
Different from those contributions, our work is interested in the impact of aged CSIT on the massive-MIMO-enabled URSST. 
Unlike the work in \cite{8648406,8836598} which consider deterministic time-varying channel models for the aged CSIT, our investigation is mainly based on the first-order Markov model of the aged CSIT, which is more appropriate for wireless environments that are rich in scattering. 
It is perhaps worth noting that MIMO (or massive-MIMO) beamforming with aged CSIT has already received intensive investigation for throughput-oriented systems and for the average performance (e.g., \cite{7120183,7307172,7248611}), while our work targets on the reliability for every single shot. 
Under this new problem context, major contributions of our work include:

{\em 1)}  We investigated the performance of superimposed MF beamforming and time-orthogonal MF beamforming with the pessimistic transmit-power adaptation. 
When the receiver has multiple receive-antennas, it is shown that the superimposed MF beamforming cannot maximize the iSNR for each individual antenna. 
For this reason, the time-orthogonal MF beamforming generally outperforms the superimposed MF beamforming.
However, the price for that is the transmission latency. 
The performance-latency tradeoff is extensively studied through computer simulations.

{\em 2)} 
The pessimistic transmit-power adaptation requires the pessimistic iSNR prediction, which is based on a good lower bound of the beamforming gain for every single shot. 
Our lower-bound analysis is novel in the sense that it is conducted particularly for the massive-MIMO-enabled URSST with the first-order Markov channel model, where those lower bounds available in the literature (e.g. \cite{8660712,9120745}) are not immediately applicable. 
Our theoretical work shows that the Chernoff lower bound has the advantage of tightness particularly for URSST systems with a large number of transmit-antennas. 
This finding motivates the use of the Chernoff lower bound for the pessimistic iSNR prediction and accordingly the pessimistic transmit-power adaptation. 

{\em 3)} The performance of MF beamforming and transmitter-power adaptation is limited by the CSIT uncertainty which grows with the age of CSIT. 
In order to mitigate the detrimental impact of the CSIT uncertainty, a combinatorial approach of the MF beamforming and grouped space-time block code (G-STBC) is proposed. 
It is shown that the combinatorial approach can largely improve the transmit-power efficiency particularly for systems with a relatively small number of transmit-antennas and higher mobility. 
However, the performance improvement becomes smaller with the increase of transmit-antennas. 
Our analysis shows that this phenomenon is mainly due to the massive-MIMO channel hardening effect, which still presents in the case of CSIT uncertainty. 
For a channel-hardened massive-MIMO system, the impact of STBC becomes very limited. 

{\em 4)} To bring our work closer to the reality, practical short-length (e.g., $128$ bits/packet) forward-error-correction (FEC) codes are considered in our numerical analysis. 
Specifically, we investigated the tail-biting convolutional code (TB-CC), low-density parity check (LDPC) codes, and extended Bose–Chaudhuri–Hocquenghem (eBCH) codes. 
It is observed that the eBCH-coded approach offers the best performance in terms of the transmit-power efficiency. 

The rest of this paper is organized as follows. Section \ref{secII}  presents the system model and problem statement. Section \ref{secIII} presents the lower bound of beamforming gain for massive-MIMO MF beamforming with aged CSIT. Section \ref{secIV} presents the combinatorial approach of MF beamforming and G-STBC. Section \ref{secV} presents numerical and simulation results, and finally Section \ref{secVI} draws the conclusion.

\subsubsection*{Notations}
Regular letter, lower-case bold letter, and capital bold letter represent scalar, vector, and matrix, respectively. 
$\Re(\cdot)$ and $\Im(\cdot)$ represent the real and imaginary parts of a complex number, respectively. 
The notations $[\cdot]^T$, $[\cdot]^H$, $[\cdot]^*$, $|\cdot|$, $\left \| \cdot \right \|$ represent the transpose, Hermitian, conjugate, modulus, Frobenius norm (or Euclidean norm) of a matrix (a vector or a scalar if appropriate), respectively. $\mathbb{E}\left [ \cdot \right ]$ denotes the expectation and $\mathbf{I}_N$ denotes the $(N)\times(N)$ identity matrix.

\section{System Model and Problem Statement}\label{secII}
\subsection{System Description and Modeling}
Consider a wireless network where a massive-MIMO access point with $M$ transmit-antennas communicates to a set of mobile terminals with each having $N$ receive-antennas ($M\gg N$).
It is assumed that the transmitter (Tx)-receiver (Rx) links {are made orthogonal through frequency-division multiple access (FDMA)} \footnote{Time-domain multiple-access (TDMA) is not preferred as they are less latency friendly.}.
This assumption helps to focus our technical discussion on the single-user massive-MIMO problem context in terms of the SNR. 
{More importantly, practical URSST (or URLLC) systems are often noise limited as they are extremely demanding to the reliability \cite{8660457}.}

For the mathematical modeling, the following systems specification applies:

\begin{enumerate}
\item[s1)] {\bf Short block-length}: this is referred to {the} short-length packet (typically $128\sim1,000$ bit/blk \cite{3GPPTR38913}), where the finite block-length communication theory applies. 
\item[s2)] {\bf Single-shot transmission}: the packet is transmitted only once. 
No retransmission is allowed or considered for the sake of latency-critical applications or tasks.
\item[s3)] {\bf Ultra-reliability}: the packet must be successfully transmitted to the Rx at an ultra-high probability (such as $99.999\%$ in the 3GPP document \cite{3GPPTR38913} or even as high as $99.99999\%$ for some industrial use cases \cite{8723572}).
It means that every Tx-Rx link should have a statistically guaranteed iSNR, e.g., $\mathscr{P}(\mathrm{isnr}\geq\mathrm{isnr}_0)\geq 99.999\%$, 
where $\mathrm{isnr}_0$ is the iSNR threshold; see Section \ref{secIII}-\ref{secV} for the detailed discussion.
\end{enumerate}
Providing this system specification, the Tx-Rx signal model has the following standard vector/matrix form \footnote{The large-scale path-loss is assumed to be constant in our work, and it is abbreviated in \eqref{eqn01} for the notation simplicity.}
\begin{equation}\label{eqn01}
\mathbf{y}=\sqrt{\gamma}\mathbf{H}\mathbf{x}+\mathbf{v},
\end{equation}
where $\mathbf{y}\in\mathbb{C}^{N\times1}$ stands for a $(N)\times(1)$ received vector, 
$\mathbf{x}\in\mathbb{C}^{M\times1}$ for an $(M)\times(1)$ transmitted vector with $\mathbb{E}(\mathbf{x}^H\mathbf{x})=1$, 
$\mathbf{H}$ for a $(N)\times (M)$ MIMO channel transition matrix,
$\mathbf{v}\sim\mathcal{CN}(\mathbf{0}, N_0\mathbf{I}_N)$ for the additive white Gaussian noise (AWGN), 
and $\gamma$ for the transmit power. 
 {It is worth noting that the MIMO channel is not considered as block fading. 
Instead, its time-varying characteristic is described by the first-order Markov model: $\mathbf{H}_\tau=\mathcal{J}_0(2\pi f_\mathrm{d})\mathbf{H}_{\tau-1}+\mathbf{\Omega}_\tau$, where $\mathcal{J}_0(\cdot)$ stands for the zero-order Bessel function of the first kind, 
$f_\mathrm{d}$ for the maximum Doppler shift, 
$\boldsymbol{\Omega}_\tau\in\mathbb{C}^{N\times M}$ for an i.i.d. complex Gaussian matrix, and $\tau$ for the time-slot index of the channel, respectively.
This channel model has been widely used to study various wireless systems in slowly or moderately time-varying channels (e.g., \cite{891214,1593619}).
Using the approximation $\mathcal{J}_0^\tau(2\pi f_\mathrm{d})\approx\mathcal{J}_0(2\pi \tau f_\mathrm{d})$ (see \cite{1593619}), $\mathbf{H}_\tau$ can be approximately represented by: 
\begin{equation}\label{eqn02}
\mathbf{H}_\tau\approx\mathcal{J}_0(2\pi f_\mathrm{d}\tau)\mathbf{H}_0+\mathbf{\Omega},
\end{equation}
where each element of $\mathbf{\Omega}$ follows the distribution $\mathcal{CN}(0, \sigma_\omega^2)$ with $\sigma_\omega^2=1-\mathcal{J}_0^2(2\pi f_\mathrm{d}\tau)$. Since the term $(2\pi f_d)$ keeps constant throughout the rest of the paper,} it is omitted in the Bessel function for the sake of notation simplicity, {i.e., $\mathcal{J}_0(\tau)\triangleq\mathcal{J}_0(2\pi f_\mathrm{d}\tau)$}.

Concerning the system specification: s1) short-length message (i.e., low data-rate transmission) and s3) ultra-reliability, the case where the Tx-Rx link has only single spatial data-stream is of particular interest. 
This is because the single-streaming approach allows the Tx-Rx link to exploit all available spatial degrees-of-freedom (DoF) for the diversity-gain (i.e., the reliability enhancement). 
With this insight in mind, we define $\mathbf{x}\triangleq \mathbf{w}s$, where $\mathbf{w}\in\mathbb{C}^{M\times 1}$ is the Tx-side beamforming vector with $\mathbf{w}^H\mathbf{w}=1$, and $s$ the information-bearing symbol with $\mathbb{E}(ss^*)=1$. 
Moreover, the Rx-side beamforming vector $\mathbf{u}\in\mathbb{C}^{N\times 1}$ is employed to combine the received vector $\mathbf{y}$ as
\begin{IEEEeqnarray}{ll}
z&=\mathbf{u}^H\mathbf{y},\label{eqn03}\\
&=\sqrt{\gamma}\mathbf{u}^H\mathbf{H}\mathbf{w}s+\mathbf{u}^H\mathbf{v}.\label{eqn04}
\end{IEEEeqnarray}
Then, the information-bearing symbol $s$ can be recovered from $z$, with the performance determined by the iSNR:
\begin{equation}\label{eqn05}
\mathrm{isnr}=\frac{\gamma|\mathbf{u}^H\mathbf{H}\mathbf{w}|^2}{\mathbf{u}^H\mathbf{u}}.
\end{equation}

\subsection{Problem Statement}
When $\mathbf{H}$ is available both at the Tx and the Rx \footnote{It is assumed that the Tx-side channel knowledge is made available either through channel feedback or taking advantage of the channel reciprocity in the time-division duplexing (TDD).}, 
we can easily manage the iSNR by manipulating the parameters $\gamma$, $\mathbf{w}$, and $\mathbf{u}$. 
{For instance, we can apply the singular-value decomposition (SVD) on $\mathbf{H}$, i.e., $\mathbf{H}=\mathbf{P}\mathbf{\Lambda}\mathbf{Q}^H$, where $\mathbf{P}$ is a $(N)\times(N)$ unitary matrix, $\mathbf{Q}^H$ the matrix formed by the first $N$ rows of an $(M)\times(M)$ unitary matrix,  
and $\mathbf{\Lambda}$ is the diagonal matrix with $N$ non-zero singular-values in its diagonal. 
For the single-streaming approach, the Tx beamforming-vector $\mathbf{w}$ and Rx beamforming-vector $\mathbf{u}$ can be formed by
\begin{equation}\label{eqn06}
\mathbf{w}=\frac{\sum_{n=0}^{N-1}\mathbf{q}_n}{N}, ~ \mathbf{u}=\sum_{n=0}^{N-1}\mathbf{p}_n,
\end{equation}
where $\mathbf{q}_n$ is the $n^\mathrm{th}$ column of $\mathbf{Q}$, and $\mathbf{p}_n$ is the $n^\mathrm{th}$ column of $\mathbf{P}$.
Applying \eqref{eqn06} into \eqref{eqn05} yields}
\begin{equation}\label{eqn07}
\mathrm{isnr}=\frac{\gamma\left(\sum_{n=0}^{N-1}\lambda_n\right)^2}{N},
\end{equation}
where $\lambda_n(>0)$ ($n=0,...,N-1$) are the singular-values of $\mathbf{H}$. 
For {practical massive-MIMO systems}, SVD is a high-complexity algorithm, and the matched-filter (MF) is often employed as a low-complexity alternative {(e.g., \cite{1564285,1468466})}.
Conventional throughput-oriented MF beamforming often superimposes multiple data-streams for better spectral efficiency.
For the case of single data-stream, the superimposed MF beamforming reduces to
\begin{equation}\label{eqn08}
\mathbf{w}=\frac{\sum_{n=0}^{N-1}\mathbf{h}_n^*}{\|\sum_{n=0}^{N-1}\mathbf{h}_n\|},~\mathbf{u}=\mathbf{H}\mathbf{w},
\end{equation}
where $\mathbf{h}_n^T$ is $n^\mathrm{th}$ row of $\mathbf{H}$.
In this case, the iSNR in \eqref{eqn07} becomes 
\begin{equation}\label{eqn09}
\mathrm{isnr}=\gamma\|\mathbf{H}\mathbf{w}\|^2.
\end{equation} 
However, it is shown in \eqref{eqn08} that the optimality of $\mathbf{w}$ is lost since $\mathbf{w}$ is no longer optimized for any of the Rx antennas.
Alternatively, the time-orthogonal MF beamforming can improve the iSNR at the cost of higher transmission latency. 
{With $N$ orthogonal time slots, the Tx beamforming aims at only the $n^\mathrm{th}$ Rx antenna in the $n^\mathrm{th}$ slot: 
\begin{equation}\label{eqn10}
\mathbf{w}_n=\left(\mathbf{h}^*_{n}\right)/\|\mathbf{h}_{n}\|,~\mathbf{u}=[\mathbf{h}_0^T\mathbf{w}_0,...,\mathbf{h}_{N-1}^T\mathbf{w}_{N-1}].
\end{equation}
Then, the iSNR in \eqref{eqn05} becomes}
\begin{equation}\label{eqn11}
\mathrm{isnr}=\gamma\|\mathbf{H}\|^2.
\end{equation}
{It is perhaps worth noting that the coding gain is not considered in the above derivation due to the use of repetition code. 
The impact of coding gain will be discussed in Section \ref{secV}.}

For the time-varying channel described in \eqref{eqn02}, 
it is not practical to assume the availability of $\mathbf{H}$ at the Tx. 
This is particularly true for a relatively high mobility, where the channel knowledge imperfection grows considerably with the time delay. 
{Assume that the Tx-side channel knowledge is $\mathbf{H}_0$}, which was obtained at the initial time interval. 
Due to the feedback and/or processing delay, $\mathbf{H}_0$ is utilized to form $\mathbf{w}$ at the time $\tau$, where the MIMO channel has already become $\mathbf{H}_\tau$.
Taking the superimposed beamforming as an example, the actual iSNR at the time interval $\tau$ becomes 
{\begin{equation}\label{eqn12}
\mathrm{isnr}=\gamma\beta_{\tau},
\end{equation}  
where $\beta_{\tau}$ is the beamforming gain given by
\begin{equation}\label{eqn13}
\beta_{\tau}\triangleq\|\mathbf{H}_\tau\mathbf{w}\|^2.
\end{equation}
Tx does not know exactly about $\beta_{\tau}$, which introduces uncertainty} to the transmit-power adaptation.
To ensure the ultra-reliability, Tx needs to predict a threshold value of $\beta_{\tau}$. 
To focus on the analysis of $\beta_{\tau}$, it is assumed that the transmission counts as failure once the outage occurs (similar assumption used in the literature, e.g., \cite{9120745,8917650}). In this case, the PER can be computed as:
\begin{IEEEeqnarray}{rl}
\mathscr{P}_\textsc{per}&\approx \mathscr{P}_\textsc{out}+(1-\mathscr{P}_\textsc{out})\mathscr{P}_\textsc{dec},\label{eqn14}\\
&\approx \mathscr{P}_\textsc{out}+\mathscr{P}_\textsc{dec},\label{eqn15}
\end{IEEEeqnarray}
where $\mathscr{P}_\textsc{per}$ stands for the PER, $\mathscr{P}_\textsc{out}$ for the outage probability, and $\mathscr{P}_\textsc{dec}$ for {the decoding error rate of a packet when the system is not outage.} 

\begin{figure}[t]
\centering
\includegraphics[scale=0.2]{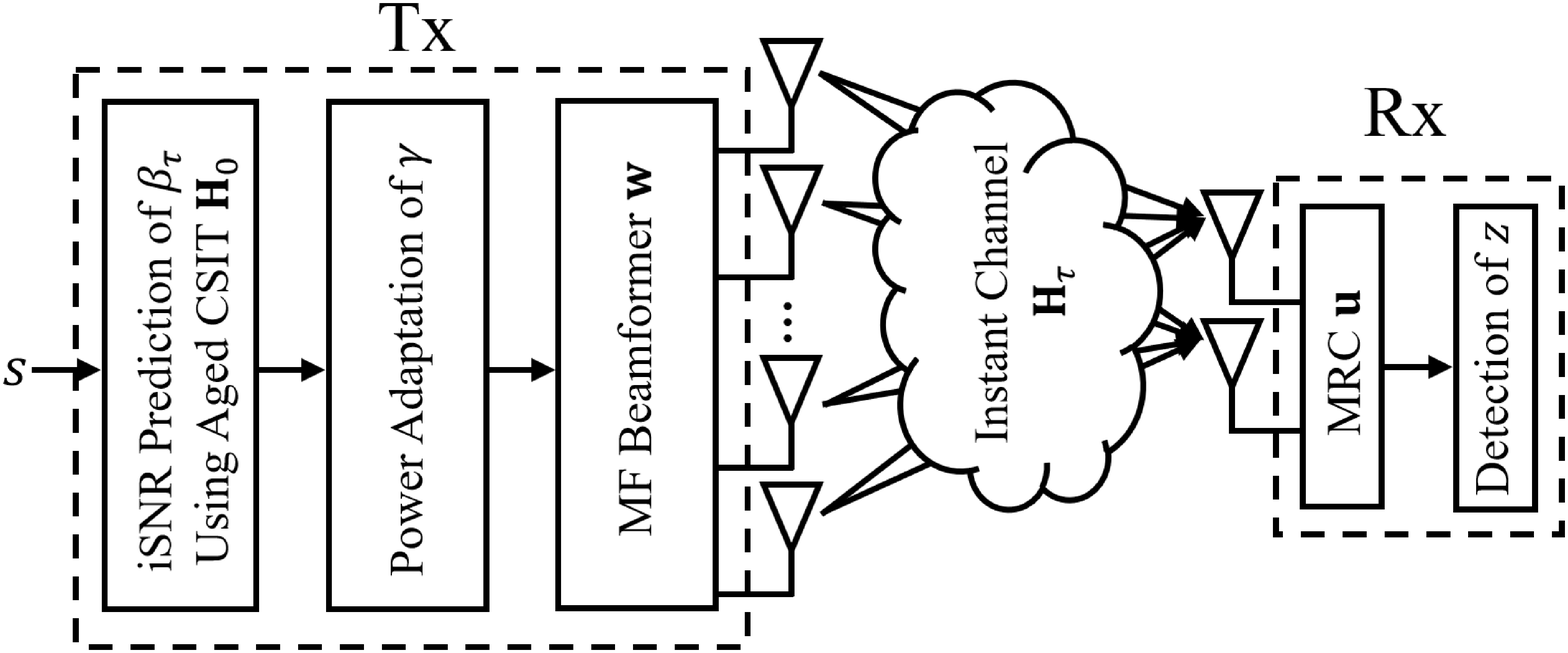}
\caption{Block diagram of the massive-MIMO MF beamforming system with pessimistic power adaptation for URSST.}
\vspace{-0em}
\label{fig0}
\end{figure}

A pessimistic condition to meet the system specification s3) is: 
\begin{equation}\label{ym016}
2\max(\mathscr{P}_\textsc{out}, \mathscr{P}_\textsc{dec})\leq\overline{\mathscr{P}_\textsc{per}},
\end{equation}
where $\overline{\mathscr{P}_\textsc{per}}$ is the target PER of the single shot. 
To ensure the target PER, we set the iSNR threshold, $\mathrm{isnr}_0$, according to \eqref{ym016} and aim to find an iSNR fulfilling 
$\mathrm{isnr}\geq\mathrm{isnr}_0$.
Using the result \eqref{eqn12}, the transmit power should fulfill 
\begin{equation}\label{eqn17}
\gamma\beta_{\tau}\geq\mathrm{isnr}_0.
\end{equation} 
As the beamforming gain, $\beta_\tau$, is unknown to the Tx,  our aim is to find a lower bound $\beta_\tau^\perp\leq\beta_\tau$ and ensure
\begin{equation}\label{eqn18}
\gamma\beta_{\tau}\geq\gamma\beta_{\tau}^{\perp}\geq\mathrm{isnr}_0,
\end{equation} 
with which the reliability requirement is satisfied.
Since $\beta_\tau$ is dependent on the specific transmission scheme, 
it will be carefully investigated for the superimposed MF beamforming, time-orthogonal MF beamforming as well as their combinations with the G-STBC.

\section{{Pessimistic Power Adaptation for MF Beamforming}}\label{secIII}
In this section, we will first study lower bounds of the MF beamforming gain in the case of aged CSIT, and then use them to conduct the pessimistic power adaptation for the MF beamforming in the URSST system. 
\subsection{Lower Bounds of The MF Beamforming Gain}
By applying \eqref{eqn02} into \eqref{eqn13}, the superimposed MF beamforming gain, $\beta_\tau$, is expanded into
\begin{equation}\label{ym019}
\beta_\tau=\mathcal{J}_0^2(\tau)\beta_0+\eta,
\end{equation}
where the first term of \eqref{ym019} is certain to the Tx, and the second term is uncertain and expressed by 
\begin{equation}\label{ym020}
\eta=2\mathcal{J}_0(\tau)\Re(\mathbf{w}^H\mathbf{H}_0^H\mathbf{\Omega}\mathbf{w})+\|\mathbf{\Omega}\mathbf{w}\|^2.
\end{equation}
It is clear that $\beta_\tau$ is a random variable obeying the non-central chi-square distribution with its mean and variance given by \cite{AP91}
\begin{equation}\label{ym021}
\mu_\text{s}(\tau)\triangleq\mathbb{E}(\beta_\tau)=\mathcal{J}_0^2(\tau)\left\|\mathbf{H}_0\mathbf{w}\right\|^2+N\sigma_\omega^2,
\end{equation}
\begin{equation}\label{ym022}
\sigma_\text{s}^2(\tau)\triangleq\mathrm{Var}(\beta_\tau)=4\mathcal{J}_0^2(\tau)\left\|\mathbf{H}_0\mathbf{w}\right\|^2+2N\sigma_\omega^2.
\end{equation}
Similarly, it is trivial to find that the time-orthogonal MF beamforming gain also obey the non-central chi-square distribution, 
and the mean and variance of which is given by
\begin{equation}\label{ym023}
\mu_\text{o}(\tau)=\mathcal{J}_0^2(\tau)\|\mathbf{H}_0\|^2+N\sigma_\omega^2,
\end{equation}
\begin{equation}\label{ym024}
\sigma_\text{o}^2(\tau)=4\mathcal{J}_0^2(\tau)\left\|\mathbf{H}_0\right\|^2+2N\sigma_\omega^2.
\end{equation}
To have unified notations for the rest of the paper, or otherwise specified, we use $\mu(\tau)$ and $\sigma^2(\tau)$ to denote the mean and variance of $\beta_\tau$.
With the above results in place, we are ready to analyze the lower bound of the MF beamforming gain, i.e., $\beta_\tau^\perp$ in \eqref{eqn18}.

\begin{lem}[Polynomial expansion lower bound from \cite{8660712}]\label{lem01}
Consider the random variable, $\beta_\tau$, in form of \eqref{ym019} and the probability $\mathrm{Pr}(\beta_\tau\geq\beta_{\tau,\text{poly}}^\perp)\leq\mathscr{P}_\textsc{out}$.
For $\mathscr{P}_\textsc{out}\to 0$, the lower bound $\beta_{\tau,\text{poly}}^\perp$ is given by
\begin{equation}\label{ym025}
\beta_{\tau, \text{poly}}^\perp=\left(\mathscr{P}_\textsc{out}N!\right)^{\frac{1}{N}}\sigma_\omega^2\exp\left(\frac{\mu(\tau)}{N\sigma_\omega^2}-1\right).
\end{equation}
\end{lem}

\begin{lem}[derived from \cite{1650344,1542405}]\label{lem02}
For the random variable, $\beta_\tau$, in form of \eqref{ym019}, its Chebyshev lower bound, $\beta_{\tau,\text{Cheby}}^\perp$, should satisfy
\begin{equation}\label{ym026}
\left|\beta_{\tau,\text{Cheby}}^\perp-\mu(\tau)\right|=\frac{\sigma(\tau)}{\sqrt{\mathscr{P}_\textsc{out}}}.
\end{equation}
\end{lem}

\begin{lem}[derived from \cite{9120745,5963622}]\label{lem03}
Define a function
\begin{equation}\label{ym027}
f(t,\beta_{\tau}^\perp)\triangleq \exp(t\beta_{\tau}^\perp)\mathbb{E}\left(\exp(-t\beta_{\tau})\right). 
\end{equation}
The Chernoff lower bound, $\beta_{\tau,\text{Cher}}^\perp$, should satisfy
\begin{equation}\label{ym028}
\min_{t>0}f(t,\beta_{\tau,\text{Cher}}^\perp)=\mathscr{P}_\textsc{out}.
\end{equation}
\end{lem}

In Section V, all of these three bounds are examined in terms of their fitness to the pessimistic power adaptation, and the Chernoff lower bound is found the most appropriate one. 
It is perhaps worth noting that the Jensen's bound (e.g. \cite{7727938,8345703}) is not suitable here since the convexity of non-central chi-square distributions varies with respect to their means and variances. 
In {\em Lemma \ref{lem03}}, \eqref{ym028} shows that the Chernoff bound does not exhibit a closed-form. 
Therefore, the following result is provided to facilitate our study. 

\begin{thm}\label{thm01}
For an arbitrarily small $\mathscr{P}_\textsc{out}$ in \eqref{ym028}, the Chernoff lower bound $\beta_{\tau,\text{Cher}}^\perp$ falls in the range of $\beta_{\tau,\text{Cher}}^\perp\in\left(0,\mu(\tau)\right)$, and $\beta_{\tau,\text{Cher}}^\perp$ is a monotonically increasing function of $\mathscr{P}_\textsc{out}$.
\end{thm}
\begin{IEEEproof}
According to the proof in Appendix \ref{appdx1}, \eqref{ym027} can be represented into the following form

\begin{equation}\label{ym029}
f(t,\beta_{\tau}^\perp)=\frac{\exp\left(t\beta_{\tau}^\perp-\frac{(\mu(\tau)-N\sigma_\omega^2)t}{1+\sigma_\omega^2t}\right)}{(1+\sigma_\omega^2t)^{N}}.
\end{equation}
Then, the partial derivative of $f(t,\beta_{\tau}^\perp)$ with respect to $t$ is computed by
\begin{IEEEeqnarray}{ll}\label{ym030}
\frac{\partial f(t,\beta_{\tau}^\perp)}{\partial t}&=f(t,\beta_{\tau}^\perp)\left(\beta_{\tau}^\perp-\frac{\mu(\tau)-N\sigma_\omega^2}{\left(1+\sigma_\omega^2t\right)^2}-\frac{N\sigma_\omega^2}{1+\sigma^2_\omega t}\right),\nonumber\\
&>f(t,\beta_{\tau}^\perp)(\beta_{\tau}^\perp-\mu(\tau)).
\end{IEEEeqnarray}
The inequality \eqref{ym030} holds due to $(1+\sigma_\omega^2t)>1$. Now, we consider two cases of $\beta_{\tau}^\perp$ as follows:

{\it Case 1:} $\beta_{\tau}^\perp\geq\mu(\tau)$. We immediately have $\partial f(t,\beta_{\tau}^\perp)/\partial t>0$ and understand that 
$f(t,\beta_{\tau}^\perp)$ is a monotonically increasing function of $t$. 
Then, the following result holds 
\begin{equation}\label{ym031}
\min_{t>0}f(t,\beta_{\tau,\text{Cher}}^\perp)=f(t\to 0,\beta_{\tau,\text{Cher}}^\perp)=1.
\end{equation}

{\it Case 2:} $0<\beta_{\tau}^\perp<\mu(\tau)$. We let $\partial f(t,\beta_{\tau}^\perp)/\partial t=0$ and obtain
\begin{equation}\label{ym032}
\beta_{\tau}^\perp-\frac{\mu(\tau)-N\sigma_\omega^2}{\left(1+\sigma_\omega^2t\right)^2}-\frac{N\sigma_\omega^2}{1+\sigma^2_\omega t}=0.
\end{equation}
Solving \eqref{ym032} results in
\begin{equation}\label{ym033}
t^\star=\frac{N\sigma_\omega^2+\sqrt{\sigma_\omega^4N^2+4\beta_{\tau}^\perp\left(\mu(\tau)-N\sigma_\omega^2\right)}}{2\sigma_\omega^2\beta_{\tau}^\perp}-\frac{1}{\sigma_\omega^2},
\end{equation}
with which $f(t,\beta_{\tau}^\perp)$ reaches the minimum. 

Considering the case of $\beta_{\tau}^\perp\to 0$, \eqref{ym033} immediately gives $t^\star\to\infty$. 
Applying this result into \eqref{ym029} gives: 
\begin{equation}
f(t^\star,\beta_{\tau}^\perp\to 0)\to0.
\end{equation}
With \eqref{ym029}, we understand that $f(t^\star,\beta_{\tau}^\perp)$ is a monotonically increasing function of $\beta_{\tau}^\perp$. 
Therefore, the minimum of $f(t^\star,\beta_{\tau}^\perp)$ increases with the increase of $\beta_{\tau}^\perp$ and so as for $\mathscr{P}_\textsc{out}$.

As a conclusion, for an arbitrarily small $\mathscr{P}_\textsc{out}$, {\it Case 1} is not a valid option as we cannot have a valid Chernoff lower bound that can satisfy the condition \eqref{ym028}. {\it Theorem \ref{thm01}} is therefore proved. 
\end{IEEEproof}

The important message from {\em Theorem \ref{thm01}} is that, for an arbitrarily small $\mathscr{P}_\textsc{out}$, the Chernoff lower bound, $\beta_{\tau,\text{Cher}}^\perp$, can be found through the line searching over the range, $\beta_{\tau}^\perp\in\left(0,\mu(\tau)\right)$.
Moreover, it can be observed that the mean, $\mu(\tau)$, is the upper bound of $\beta_{\tau,\text{Cher}}^\perp$. 

\subsection{The Hardening Effect of MF Beamforming Gain}
With the perfect CSIT, MF beamforming enjoys the hardened iSNR for $M\to\infty$ (e.g., \cite{7827017,6799319}). 
Therefore, it would be interesting to study whether such hardening effect still holds in the case of aged CSIT.  
To prevent the power growing with the increase of antennas, the beamforming gain $\beta_\tau$ is normalized by the factor $(MN)$.
\begin{cor}\label{cor01}
Suppose $M\rightarrow\infty$, the following convergence holds for $\beta_{\tau}^\perp$:	
\begin{equation}\label{ym035}
\lim\limits_{M\rightarrow\infty}\beta_{\tau}^\perp=\lim\limits_{M\rightarrow\infty}\mu(\tau).
\end{equation}
\end{cor}
\begin{IEEEproof}
Given the optimal configuration $t^\star$, \eqref{ym032} shows that $\beta_{\tau}^{\perp}$ after normalization reads as
\begin{equation}\label{ym036}
\beta_{\tau}^\perp=\frac{\mu(\tau)-(\sigma_\omega^2)/(M)}{\left(1+(\sigma_\omega^2t^\star)/(MN)\right)^2}+\frac{N\sigma_\omega^2}{MN+\sigma_\omega^2t^\star}.
\end{equation}	
For $M\rightarrow\infty$, it is trivial to confirm the result \eqref{ym035}. {\it Corollary \ref{cor01}} is therefore proved.
\end{IEEEproof}

The convergence behaviour shown in \textit{Corollary \ref{cor01}} is reasonable since the channel uncertainty gradually vanishes with $M\rightarrow\infty$.
However, the exact form of $\lim\limits_{M\rightarrow\infty}\mu(\tau)$ is dependent on the channel fading behavior. 
Given the popularity of Rayleigh fading channels in wireless communications (e.g., \cite{9120745,9107489}), {\em Corollary \ref{cor02}} presents the exact form for this particular case.
\begin{cor}\label{cor02}
Suppose the channel to be Rayleigh fading, the following exact form holds for the superimposed and time-orthogonal beamforming, respectively:
\begin{equation}\label{ym037}
\lim\limits_{M\rightarrow\infty}\mu_\text{s}(\tau)=\frac{\mathcal{J}_0^2(\tau)}{N},~\lim\limits_{M\rightarrow\infty}\mu_\text{o}(\tau)=\mathcal{J}_0^2(\tau).
\end{equation} 
\end{cor}
\begin{proof}
See Appendix \ref{appdx2}.
\end{proof}

\textit{Corollary \ref{cor02}} also shows that the time-orthogonal beamforming gain is much higher than the superimposed beamforming gain for $M\to\infty$. This is not surprising because the superimposed beamforming trades off its performance for the latency.
This issue will be further discussed in Section \ref{sec33}.

\subsection{Pessimistic Power Adaptation}\label{sec33}
In summary, the pessimistic power adaptation algorithm includes four basic steps:

{\it Step 1:} Given a practical error-correction coding scheme, determine the iSNR threshold, $\mathrm{isnr}_0$, 
which fulfills the pessimistic condition \eqref{ym016};

{\it Step 2:}  Use the CSIT, $\mathbf{H}_0$, to form the MF beamforming vector or matrix if appropriate;

{\it Step 3:} With \eqref{ym029} and \eqref{ym033}, use the line searching algorithm to find the Chernoff lower bound $\beta_{\tau,\text{Cher}}^\perp$;

{\it Step 4:}  Use \eqref{eqn18} to determine the transmit power, $\gamma$. 

Implicitly, the pessimistic power adaptation algorithm assumes the knowledge of $\mathcal{J}_0(\tau)$, 
which varies with respect to the lag $\tau$ and the maximum Doppler shift $f_\mathrm{d}$. 
We argue that these two parameters are not hard to obtain.
In case of insufficiently accurate estimation of the velocity, a pessimistic estimate can be used instead for the sake of reliability. 
Moreover, when the number of transmit-antennas becomes sufficiently large, 
the Tx can directly use the hardened beamforming gain in \eqref{ym037} to determine the transmit power. 

It is clear that the transmit power is inversely proportional to $\mathcal{J}^2_0(\tau)$. 
Denote $\overline{\gamma}$ to be the cap of the transmitter power. 
The age of CSIT (i.e., the lag $\tau$) is capped by the factor $(\mathrm{isnr}_0)/(\overline{\gamma})$.
More specifically, for the time-orthogonal MF beamforming, the lag (normalized by $f_\text{d}$) is upper bounded by
\begin{equation}\label{ym038}
\tau f_\text{d}\leq\frac{1}{2\pi}\mathcal{J}_0^{-1}\left(\sqrt{(\mathrm{isnr}_0)/(\overline{\gamma})}\right).
\end{equation}
Similarly, for the superimposed MF beamforming, it is upper bounded by
\begin{equation}\label{ym039}
\tau f_\text{d}\leq\frac{1}{2\pi}\mathcal{J}_0^{-1}\left(\sqrt{(N)(\mathrm{isnr}_0)/(\overline{\gamma})}\right).
\end{equation}

\subsection{Superimposed Beamforming vs. Time-orthogonal Beamforming}\label{sec34}
The result \eqref{ym038}-\eqref{ym039} shows that the superimposed beamforming has a different upper bound of the CSIT age from the time-orthogonal beamforming. The latter has a much relaxed condition.
On the other hand, the latency of time-orthogonal beamforming is $N$ times of the superimposed beamforming. 
In addition, the following result shows that the time-orthogonal beamforming always has a higher beamforming gain than the superimposed beamforming. 
\begin{cor}\label{cor03}
Given the aged CSIT, $\mathbf{H}_0$, and the outage probability, $\mathscr{P}_\textsc{out}$, the Chernoff bound of the superimposed and time-orthogonal beamforming satisfies:
\begin{equation}\label{ym040}
\beta_{\tau,\text{o}}^\perp\geq\beta_{\tau,\text{s}}^\perp.
\end{equation}	
\end{cor}
\begin{IEEEproof}
With \eqref{ym021} and \eqref{ym023}, it is trivial to prove the relationship: $\mu_\text{o}(\tau)\geq\mu_\text{s}(\tau)$.
Applying this result into \eqref{ym029} leads to
\begin{equation}\label{ym041}
\mathscr{P}_\textsc{out}=f(t^\star_\text{s},\beta_{\tau,\text{s}}^\perp,\mu_\text{s}(\tau))\geq f(t^\star_\text{s},\beta_{\tau,\text{s}}^\perp,\mu_\text{o}(\tau)).
\end{equation}	
Remind that $f(t,\beta_{\tau}^\perp,\mu(\tau))$ decreases with the decrease of $\beta_{\tau}^\perp$.
Suppose $\beta_{\tau,\text{o}}^\perp<\beta_{\tau,\text{s}}^\perp$, we should have
\begin{equation}\label{ym042}
\mathscr{P}_\textsc{out}\geq f(t^\star_\text{s},\beta_{\tau,\text{s}}^\perp,\mu_\text{o}(\tau))> f(t^\star_\text{s},\beta_{\tau,\text{o}}^\perp,\mu_\text{o}(\tau)).
\end{equation}	
Since $f(t,\beta_{\tau,\text{o}}^\perp,\mu_\text{o}(\tau))$ achieves the minimum at $t=t^\star_\mathrm{o}$, we shall have
\begin{equation}\label{ym043}
\mathscr{P}_\textsc{out}>f(t^\star_\text{s},\beta_{\tau,\text{o}}^\perp,\mu_\text{o}(\tau))\geq f(t^\star_\text{o},\beta_{\tau,\text{o}}^\perp,\mu_\text{o}(\tau))=\mathscr{P}_\textsc{out}.
\end{equation}	
\eqref{ym043} is logical contradictory. 
Hence, the hypothesis $\beta_{\tau,\text{o}}^\perp<\beta_{\tau,\text{s}}^\perp$ does not hold and \textit{Corollary \ref{cor03}} is proved.
\end{IEEEproof}

For the time-orthogonal beamforming, the Tx beams the signal to a particular Rx-antenna at a time slot, and this is conducted in turn for all Rx-antennas. 
So far, the Rx-side signal combining only cares about those targeted Rx-antennas at each time slot. 
However, other Rx-antennas might still be able to receive the signal due to energy leaking. 
Indeed, the energy leaking tends to zero when $M\to\infty$.  
However, for not-too-large massive-MIMO, the energy leaking presents and can be potentially recycled through the Rx-side signal combining.
We use the maximum-ratio combining (MRC) for the signal recycling. 
In this case, the time-orthogonal beamforming gain, $\beta_{\tau}$, becomes
\begin{equation}\label{ym044}
\beta_{\tau}=\left\|\mathbf{H}_\tau\mathbf{W}\right\|^2,
\end{equation}
where $\mathbf{W}\triangleq[\mathbf{w}_0,...,\mathbf{w}_{N-1}]$, with each column vector in $\mathbf{W}$ defined in \eqref{eqn10}. 
Accordingly, we can easily compute
\begin{equation}\label{ym045}
\mu_\text{o}(\tau)=\mathcal{J}_0^2(\tau)\left\|\mathbf{H}_0\mathbf{W}\right\|^2+N^2\sigma_\omega^2,
\end{equation}
\begin{equation}\label{ym046}
f(t,\beta_{\tau,\text{recyc}}^\perp)=\frac{\exp\left(t\beta_{\tau}^\perp-\frac{(\mu(\tau)-N^2\sigma_\omega^2)t}{1+\sigma_\omega^2t}\right)}{(1+\sigma_\omega^2t)^{N^2}}.
\end{equation}
It is perhaps worth noting that \eqref{ym045} would not change the basic principle of the Chernoff lower bound stated in {\it Theorem \ref{thm01}}.

The effect of energy harvesting on $\beta_{\tau}^\perp$ is mathematically intractable. In order to see the effect of energy recycling, we define the ratio of the aSNR for cases with or without the energy recycling as a comparison:
\begin{equation}\label{ym047}
\rho\triangleq\frac{\mathbb{E}\left(\left\|\mathbf{H}_{\tau}\mathbf{W}\right\|^2 \right)}{\mathbb{E}\left( \sum_{n=0}^{N-1}\left\|\mathbf{h}_{n,\tau}^T\mathbf{w}_{n}\right\|^2  \right)}.
\end{equation}
Specifically, for the Rayleigh fading channel, we can obtain (see Appendix \ref{appdx3} for the proof)
\begin{equation}\label{eqn45}
\rho=1+\frac{(N-1)}{M\mathcal{J}_0^2(\tau)+\sigma_\omega^2}.
\end{equation}
The effect of energy recycling in the time-orthogonal beamforming will be further studied in Section \ref{secV}.

\section{The Combinatorial Approach of MF Beamforming and G-STBC}\label{secIV}
Section \ref{secIII} shows that $\beta_{\tau}^{\perp}$ is largely dependent on $\mathcal{J}_0(\tau)$.
This motivates us to further improve $\beta_{\tau}^{\perp}$ for the circumstance when $\mathcal{J}_0(\tau)$ is low.
The idea is to combine the STBC with the MF beamforming to further improve the robustness of our approach to the CSIT uncertainty. 
It is well known that the STBC trades off the latency for the reliability \cite{771146}.  
In order to reduce the latency arising from the STBC, we propose to divide Tx-antennas into a set of groups and then apply the combinatorial approach.
Different from other antenna grouping approaches in the literature (e.g., \cite{761255,4359547}) which aim to reduce the space-time decoding complexity, 
our approach aims to mitigate the detrimental impact of the CSIT uncertainty.

Consider the case where Tx-antennas are divided into a set of groups ($K$). 
The channel from the $k^\mathrm{th}$ group ($k=0,...,K-1$) to the $n^\mathrm{th}$ Rx-antenna at the lag $\tau$ is denoted by $\mathbf{h}_{n,k,\tau}^{T}$.
Each Tx-antenna group transmits a single data-stream, and the MF beamforming is applied within the group. Taking the superimposed beamforming as an example, the beamforming vector for the $k^\mathrm{th}$ group is given by
\begin{equation}\label{eqn49}
\mathbf{w}_{k}=\frac{\sum_{n=0}^{N-1}\mathbf{h}_{n,k,0}^{*}}{\|\sum_{n=0}^{N-1}\mathbf{h}_{n,k,0}\|}.
\end{equation}
Then, each Tx-antenna group is regarded as a Tx-antenna in the conventional STBC scheme. 
This is the concept of G-STBC defined in this paper.
The signal model of MF-beamforming and G-STBC combinatorial approach can be represented by the following matrix form
\begin{equation}\label{eqn50}
\mathbf{Y}=\sqrt{\gamma}\widetilde{\mathbf{H}}_{\tau}\mathbf{X}+\mathbf{V},
\end{equation}
where $\widetilde{\mathbf{H}}_{\tau}\in\mathbb{C}^{N\times K}$ stands for the $(N)\times(K)$ equivalent channel with each element given by $\widetilde{h}_{n,k,\tau}=\mathbf{h}_{n,k,\tau}^{T}\mathbf{w}_{k}$, $\mathbf{X}$ and $\mathbf{Y}$ for the transmitted signal and received signal matrix, respectively, with their sizes determined by $N$ and $K$, and $\mathbf{V}$ for the corresponding AWGN matrix.
At the receiver, received signals are combined in the same way as that used in the conventional STBC (see Section III-C in \cite{771146}), with which the beamforming gain, $\beta_{\tau}$, becomes
\begin{equation}\label{eqn51}
\beta_{\tau}=\big\|\widetilde{\mathbf{H}}_{\tau}\big\|^2.
\end{equation}
Accordingly, we can easily compute
\begin{equation}\label{eqn52}
\mu(\tau)=\mathcal{J}_0^2(\tau)\big\|\widetilde{\mathbf{H}}_0\big\|^2+NK\sigma_\omega^2,
\end{equation}
\begin{equation}\label{eqn53}
f(t,\beta_{\tau}^\perp)=\frac{\exp\left(t\beta_{\tau}^\perp-\frac{(\mu(\tau)-NK\sigma_\omega^2)t}{1+\sigma_\omega^2t}\right)}{(1+\sigma_\omega^2t)^{NK}}.
\end{equation}
We note that the mathematical form of \eqref{eqn52} and \eqref{eqn53} also holds for the time-orthogonal beamforming. 
There is only a minor difference which applies to each element of $\widetilde{\mathbf{H}}_0$, i.e., for the time-orthogonal beamforming, we have $\widetilde{h}_{n,k,0}=\mathbf{h}_{n,k,0}^{T}\mathbf{w}_{n,k}$, where $\mathbf{w}_{n,k}$ is defined by
\begin{equation}\label{eqn54}
\mathbf{w}_{n,k}\triangleq\left(\mathbf{h}^*_{n,k,0}\right)/\|\mathbf{h}_{n,k,0}\|.
\end{equation}
Since the result \eqref{eqn52}-\eqref{eqn53} does not influence the optimization problem addressed in \textit{Theorem \ref{thm01}}, it would not change the basic principle of the Chernoff lower bound.

Another important issue is to find the grouping method that maximizes $\beta_{\tau}^\perp$. It has been shown in the proof of \textit{Corollary \ref{cor03}} that $\beta_{\tau}^\perp$ increases with the increase of $\mu(\tau)$. Hence, the following optimization problem is formulated
\begin{equation}\label{eqn55}
\max_{\mathbf{h}_{n,k,0}}~\mu(\tau),~ \mathbf{h}_{n,k,0}\subset\left\{\mathbf{h}_{n,\tau}=\bigcup_{k=0}^{K-1}\mathbf{h}_{n,k,0}\right\},~_{k=0,...,K-1}.
\end{equation}

For the superimposed beamforming, $\mu_\text{s}(\tau)$ is dependent on the way of antenna grouping. 
More specifically, the mean in \eqref{eqn52} becomes
\begin{equation}\label{eqn56}
\mu_\text{s}(\tau)=\mathcal{J}_0^2(\tau)\sum_{n=0}^{N-1}\sum_{k=0}^{K-1}\left\|\mathbf{h}_{n,k,0}^{T}\mathbf{w}_{k}\right\|^2+NK\sigma_\omega^2.
\end{equation}
In this case, \eqref{eqn55} is an integer linear programming problem, which is NP-hard. 
The optimal solution can be found through the brute force algorithm, which is however computationally prohibitive and not suitable for URSST applications.
As a low-complexity alternative, we simply group Tx-antennas according to their physical places, where a group only includes physically adjacent Tx-antennas. Without loss of generality, we assume that $M$ can be divided by $K$, with each group having $(M)/(K)$ Tx-antennas. In case there exists a reminder, we can randomly pick up $(M-K\lfloor(M)/(K)\rfloor)$ groups and increase their Tx-antennas to $\lceil(M)/(K)\rceil$.

For the time-orthogonal beamforming, the case is different. With \eqref{eqn54}, it is easy to compute
\begin{equation}\label{eqn59}
\mu_\text{o}(\tau)=\mathcal{J}_0^2(\tau)\left\|\mathbf{H}_0\right\|^2+NK\sigma_\omega^2.
\end{equation}
It can be observed that $\mu_\text{o}(\tau)$ is independent of the way of antenna grouping. 
Hence, Tx-antennas can be grouped in an ad-hoc manner, which would not influence the optimality of \eqref{eqn55}.

The proposed combinatorial approach can mitigate the detrimental effect of the CSIT uncertainty. This is because signals from different antenna groups are combined using the STBC decoder with the CSI at the receiver (i.e., $\mathbf{H}_\tau$). 
For the time-orthogonal beamforming, if we consider two grouping numbers, $K_1$ and $K_2$ ($K_1<K_2$). With \eqref{eqn53} and \eqref{eqn59}, it is rather straightforward to understand
\begin{equation}
\mathscr{P}_\textsc{out}=f(t^\star_{K_1},\beta_{\tau,K_1}^\perp,K_1)>f(t^\star_{K_1},\beta_{\tau,K_1}^\perp,K_2).
\end{equation}
Similar to \textit{Corollary \ref{cor03}}, we can prove
\begin{equation}\label{eqn61}
\beta_{\tau,K_2}^\perp>\beta_{\tau,K_1}^\perp.
\end{equation} 
Implicitly, \eqref{eqn61} reveals the trade-off between the beamforming gain and the latency, i.e., a larger number of groups will result in a higher beamforming gain while also introduce higher latency.
In the extreme case, when $K=M$, the beamforming gain of the proposed approach reduces to the same of the conventional STBC (i.e., $\beta_{\tau}^\perp=\left\|\mathbf{H}_\tau\right\|^2$).
It is worth noting that, for the superimposed beamforming, we are not able to conduct a similar analysis as it is mathematically intractable. 

With \eqref{eqn56} and \eqref{eqn59}, it is trivial to have $\mu_\text{o}(\tau)\geq\mu_\text{s}(\tau)$, and thus \textit{Corollary \ref{cor03}} still holds.
However, the latency of the time-orthogonal beamforming increases almost linearly with the factor ($NK$). 
Moreover, it will be experimentally shown in Section \ref{secV} that the performance gap between the superimposed and time-orthogonal beamforming will decrease with the increase of $K$.
This is because the antenna number in each group is decreasing, and the channel within each group is less likely to be orthogonal (i.e., avoiding the case in \textit{Corollary \ref{cor02}}).

\section{Numerical Results and Discussion}\label{secV}
In this section, extensive numerical results are presented to elaborate the theoretical work in Section \ref{secIII} and \ref{secIV}. Based on our theoretical discussions, we divide the numerical results into three examples. In the first example, we aim to demonstrate the tightness and the stochastic properties (e.g., channel hardening effect) of the Chernoff lower bound in terms of the outage probability and the PDF. In addition, the Chernoff lower bound will be compared to other approximations or mathematical bounds to show its advantage of tightness. In the second example, we aim to demonstrate the advantage of exploiting the Tx spatial diversity compared to adopting the MRC in terms of the average transmission power. In the third example, we aim to demonstrate the improvement on $\beta_{\tau}^{\perp}$ brought by the combinatorial approach of the MF beamforming and the G-STBC in terms of the average transmission power. MATLAB is used to conduct Monte-Carlo trials to study the properties of interest. The channel is assumed to be Rayleigh fading in appreciation of its popularity in the wireless communication design (e.g., \cite{9120745,9107489}). 
The central carrier frequency is assumed to be $3.5$ GHz (see \cite{Euro5GObservatory}) and the PER requirement $\mathscr{P}_\textsc{per}$ is set to be $1\times10^{-5}$ \cite{3GPPTR38913}. 
Considering the stringent latency requirement of URSST, the time lag $\tau$ is set to be $0.5$ ms.
For the pedestrian device, the user velocity is assumed to be $5$ m/s. While for the vehicular device, the user velocity is assumed to be $15$ m/s (i.e., $54$ km/h), which is typical vehicular velocity in the urban area.

\begin{figure}[t]
\centering
\begin{minipage}[t]{0.48\textwidth}
\captionsetup{width=0.96\textwidth}
\centering
\includegraphics[scale=0.29]{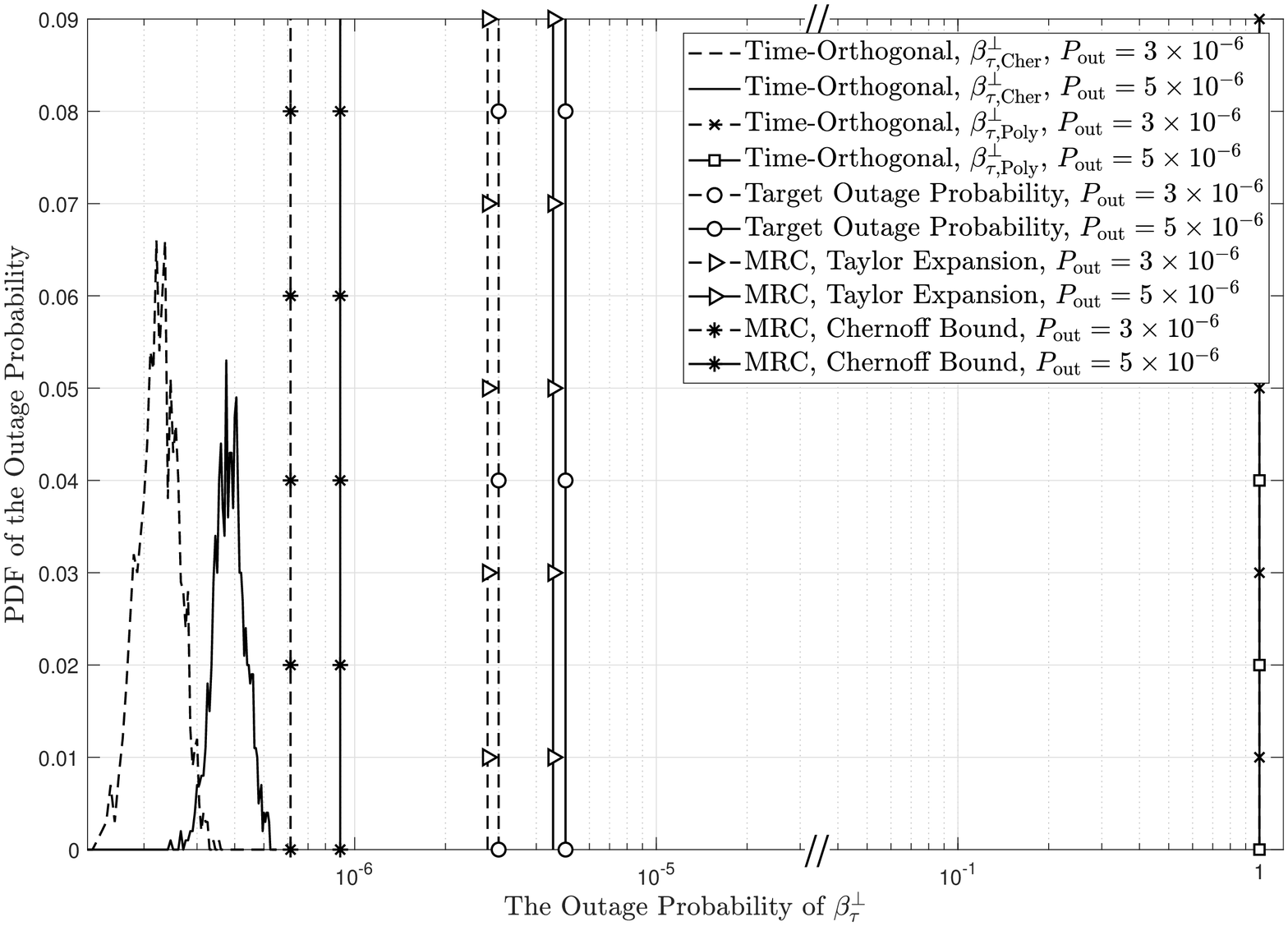}
\caption{Outage probability of $\beta_{\tau}^{\perp}$ for the time-orthogonal MF beamforming and the MRC compared to $\mathscr{P}_\textsc{out}$ when the user velocity is $15$ m/s, $M=10$, and $N=4$.}
\vspace{-0em}
\label{fig2}
\end{minipage}
\begin{minipage}[t]{0.48\textwidth}
\captionsetup{width=0.96\textwidth}
\centering
\includegraphics[scale=0.29]{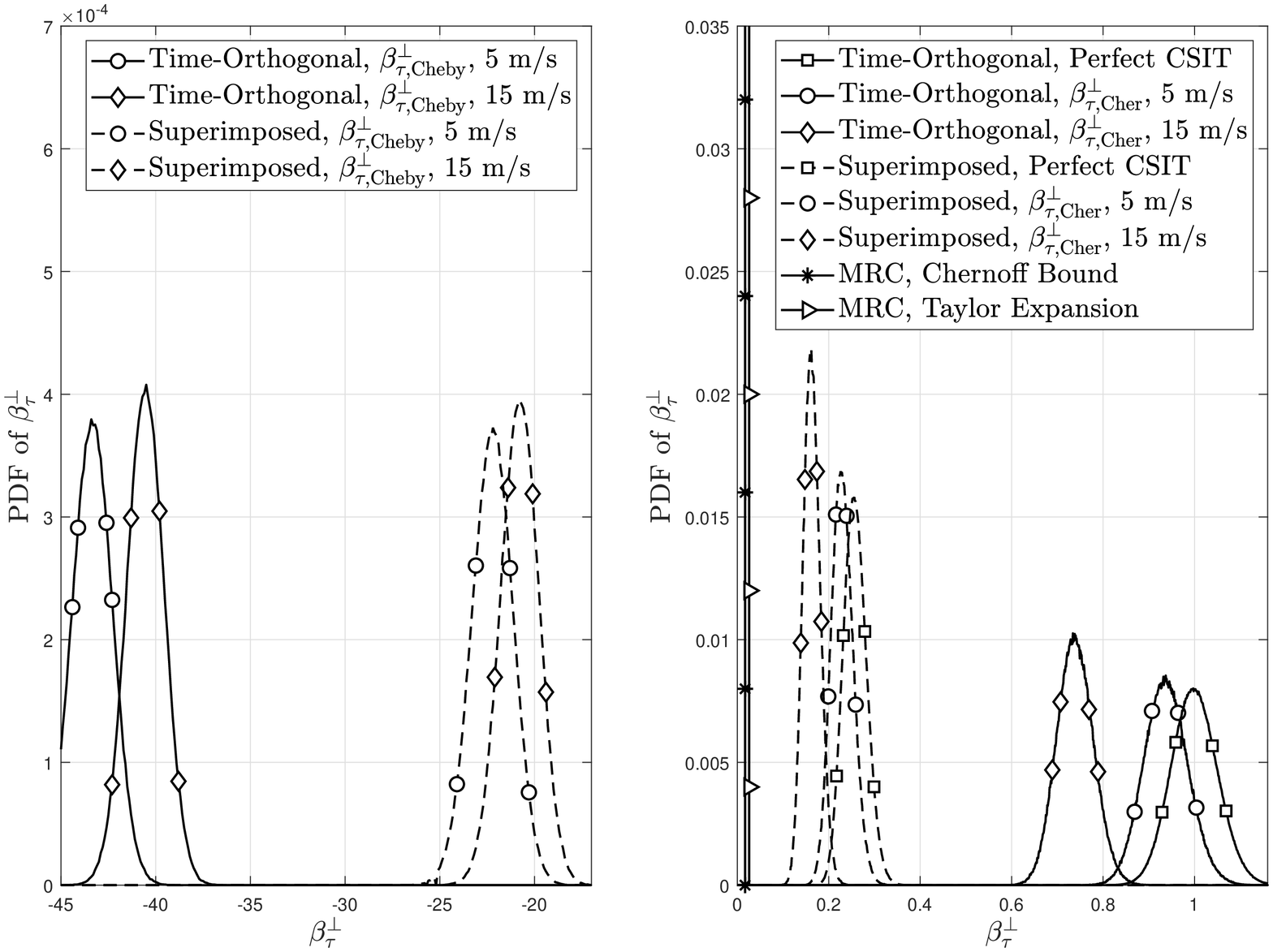}
\caption{PDF of $\beta_{\tau}^{\perp}$ for both superimposed and time-orthogonal beamforming as the user velocity changes when $M=100$, $N=4$, and $\mathscr{P}_\textsc{out}=5\times10^{-6}$.}
\vspace{-0em}
\label{fig3}
\vspace{-0em}
\end{minipage}
\end{figure}

\textit{Numerical Example 1:} This example aims to demonstrate the tightness and stochastic properties of the Chernoff lower bound. The tightness of the Chernoff lower bound is firstly demonstrated in terms of its outage probability compared to the requirement $\textsc{P}_{out}$, as shown in Fig. \ref{fig2}. 
The baselines when adopting the MRC are also included here to justify their feasibility for URSST (i.e., the Taylor expansion \cite{8660712} and the Chernoff lower bound \cite{9120745}). 
Since the Chebyshev bound is negative here and is therefore not valid to judge the outage probability, it is not included in Fig. \ref{fig2}. It is worth clarifying that specifically in Fig. \ref{fig2}, the MIMO size is set to be $M=10$ and $N=4$, which is not massive-MIMO. This is to reduce the memory size to accelerate the computing in parallel computing units when the outage probability is extremely low and will not affect the observed phenomenon. The case that the user velocity is $15$ m/s for the time-orthogonal beamforming is chosen as representative. For other user velocity or beamforming method, the phenomenon is similar. 
It can be observed that when using the MF beamforming, the Chernoff lower bound can always satisfy the outage probability requirement when $\mathscr{P}_\textsc{out}$ is either $5\times10^{-6}$ or $3\times10^{-6}$. On the other hand, when using the polynomial expansion, the outage probability is always $1$. 
This is because $\beta_{\tau, \text{poly}}^\perp$ increases exponentially with the increase of $\mu_\tau$, as shown in \eqref{ym025}. Specifically for Fig. \ref{fig2}, $\beta_{\tau, \text{poly}}^\perp$ is $10^{10}$ or higher. For massive-MIMO, this effect would be more severe. When adopting the MRC, both the Taylor expansion and the Chernoff lower bound can satisfy the outage probability requirement, and the Taylor expansion is tighter than the Chernoff lower bound.

The PDF of $\beta_{\tau}^{\perp}$ when the user velocity is changing is then shown in Fig. \ref{fig3}, where the case $\mathscr{P}_\textsc{out}=5\times10^{-6}$ is chosen as representative here. 
The polynomial expansion is not included in Fig. \ref{fig3}, since it could not fulfill the outage probability requirement as shown in Fig. \ref{fig2}. It can be observed that when using the MF beamforming, the PDF of the Chernoff lower bound moves to the left as the user velocity increases. This is reasonable, since the efficiency of exploiting the Tx spatial diversity is decreasing as $\mathcal{J}_0(\tau)$ decreases. For the time-orthogonal beamforming, $\beta_{\tau}^{\perp}$ can significantly outperform the superimposed beamforming, which confirms the conclusion in \textit{Corollary \ref{cor03}}. For the Chebyshev lower bound, it is shown that $\beta_{\tau,\text{Cheby}}^{\perp}$ is constantly negative. This is because the Chebyshev lower bound aims to find a bound of absolute value from $\mu(\tau)$. However, the PDF of $\beta_{\tau}$ is non-symmetric, and $\beta_{\tau,\text{Cheby}}^{\perp}$ becomes negative when $\mathscr{P}_\textsc{out}$ is extremely low. When adopting the MRC, $\beta_{\tau}^{\perp}$ is significantly smaller than when adopting the MF beamforming. This is reasonable, since the MRC can only exploit the Rx spatial diversity gain.


\begin{figure}[t]
\centering
\begin{minipage}[t]{0.48\textwidth}
\captionsetup{width=0.96\textwidth}
\centering
\includegraphics[scale=0.29]{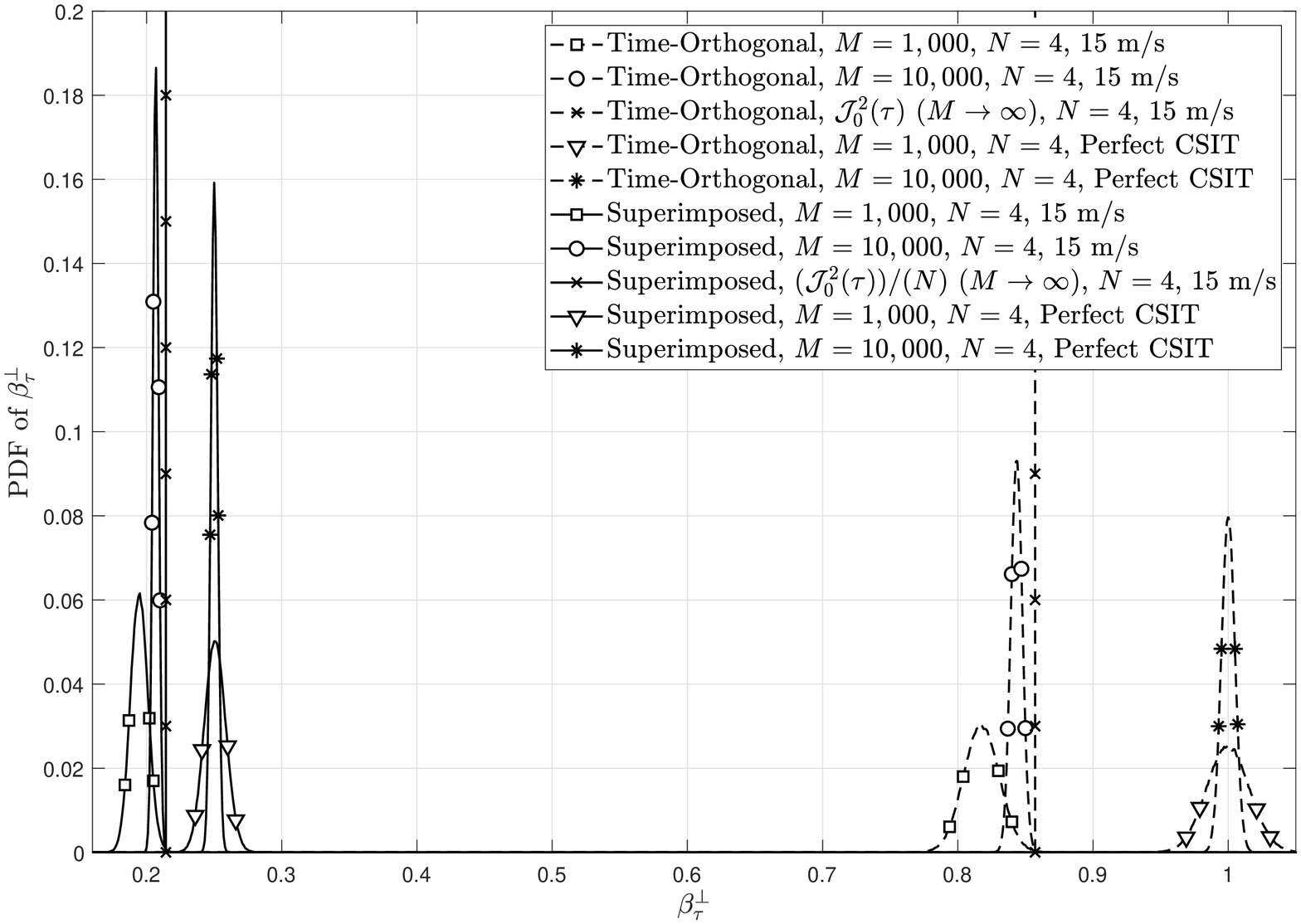}
\caption{The channel hardening effect of the Chernoff lower bound for the superimposed and time-orthogonal beamforming as the Tx antenna number $M$ increases when the user velocity is $15$ m/s, $\mathscr{P}_\textsc{out}=5\times10^{-6}$ and $N=4$.}
\vspace{-0em}
\label{fig4}
\end{minipage}
\begin{minipage}[t]{0.48\textwidth}
\captionsetup{width=0.96\textwidth}
\centering
\includegraphics[scale=0.29]{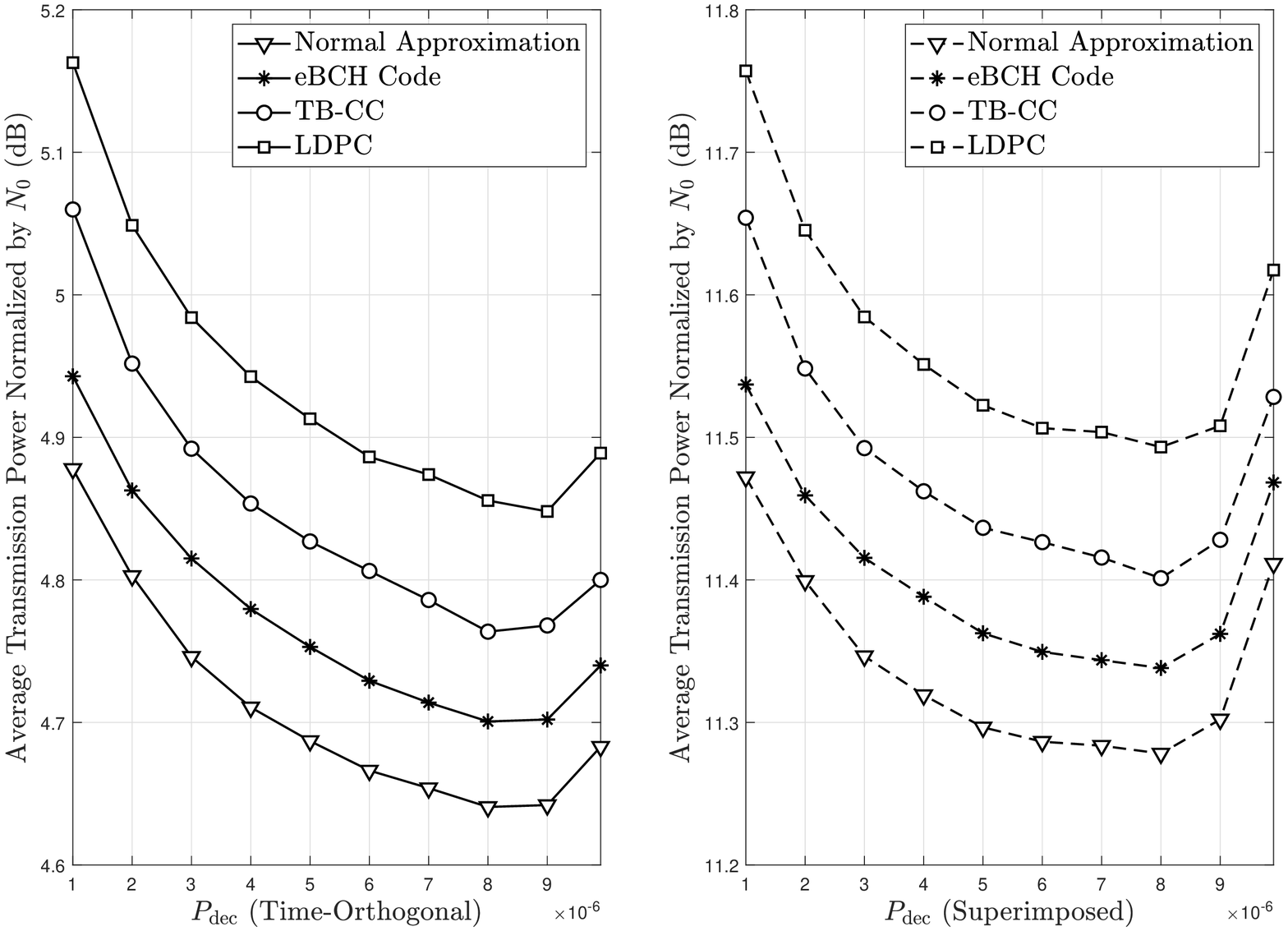}
\caption{{\footnotesize The average transmission power as $\mathscr{P}_\textsc{dec}$ is increasing for both time-orthogonal and superimposed beamforming when the user velocity is $15$ m/s, $M=100$, and $N=4$.}}
\vspace{-0em}
\label{fig5}
\end{minipage}
\end{figure}


The channel hardening effect of the Chernoff lower bound is shown in Fig. \ref{fig4}. Similar to the previous discussions, the case where the user velocity is $15$ m/s and $\mathscr{P}_\textsc{out}=5\times10^{-6}$ is chosen as representative here. It can be observed that for the time-orthogonal beamforming, $\beta_{\tau}^{\perp}$ converges to $\mathcal{J}_0^2(\tau)$. While for the superimposed beamforming, $\beta_{\tau}^{\perp}$ converges to $(\mathcal{J}_0^2(\tau))/(N)$. This coincides with \textit{Corollary \ref{cor02}}. 
Moreover, different from the channel hardening effect with perfect CSIT, it is shown that the PDF of $\beta_{\tau}^{\perp}$ moves to the right as $M$ increases. This is because the Chernoff lower bound aims to find a pessimistic bound for $\beta_{\tau}$ regarding the CSIT uncertainty, which decreases with the increase of $M$. This indicates that the spatial diversity gain is even more important for URSST when the CSIT is imperfect.

\textit{Numerical Example 2:} 
The aim of this example is to demonstrate the advantage of exploiting the Tx spatial diversity. 
To apply the power adaptation, the value of $\mathrm{isnr}_0$ needs to be determined first. 
It is perhaps worth noting that the condition in \eqref{ym016} is a little too strict. In this example, a more relaxed case is considered, where $\mathrm{isnr}_0$ is firstly determined to satisfy the decoding probability requirement $\mathscr{P}_\textsc{dec}$. Then, $\mathscr{P}_\textsc{out}$ is determined according to the approximation in \eqref{eqn15} (i.e., $\mathscr{P}_\textsc{per}\approx\mathscr{P}_\textsc{out}+\mathscr{P}_\textsc{dec}$). Apart from the widely used normal approximation, several practical FEC codes are also considered in appreciation of their good performance in the short block-length, including the TB-CC, LDPC as well as the eBCH code \cite{8594709,8815549}. The modulation is assumed to be BPSK, the codeword length is assumed to be $128$-bit, and the maximum-likelihood detection is adopted at the Rx. In this case, $\mathrm{isnr}_0$ is a monotonically decreasing function of $\mathscr{P}_\textsc{out}$ (see Fig. 3 in \cite{8594709}). 

Fig. \ref{fig5} shows the the average transmission power as $\mathscr{P}_\textsc{dec}$ increases. The case where the user velocity is $15$ m/s is chosen as representative. For the case of $5$ m/s, the phenomenon is similar. It can be observed that the average transmission power of FEC codes is only less than $0.3$ dB worse than the normal approximation. the eBCH code achieves the best performance, which is only $0.1$ dB worse than the normal approximation. More importantly, a minimum value of the average transmission power exists for each of the lines. 
This is reasonable, since when $\mathscr{P}_\textsc{dec}\to0$, $\mathrm{isnr}_0$ will be huge, and the transmission power will increase. On the other hand, when $\mathscr{P}_\textsc{dec}\to\mathscr{P}_\textsc{per}$, we have $\mathscr{P}_\textsc{out}\to0$. In this case, $\beta_{\tau}^\perp$ will decrease to almost $0$, as was specified in \textit{Theorem \ref{thm01}}. This will also increase the transmission power. For both superimposed and time-orthogonal beamforming, the average transmission power achieves the minimum value when $\mathscr{P}_\textsc{dec}=8\times10^{-6}$ in most of the cases (when the user velocity is $5$ m/s, this conclusion also holds). To minimize the power consumption, $\mathscr{P}_\textsc{dec}$ is set to be $8\times10^{-6}$ for the rest of the paper (i.e., $\mathscr{P}_\textsc{out}=2\times10^{-6}$). Moreover, using the FEC codes will only result a different value of $\mathrm{isnr}_0$, which will not affect the demonstration of exploiting the spatial diversity gain. Hence, only the normal approximation is chosen as representative in the rest of the paper.

\begin{figure}[t]
\centering
\begin{minipage}[t]{0.48\textwidth}
\captionsetup{width=0.96\textwidth}
\centering
\includegraphics[scale=0.29]{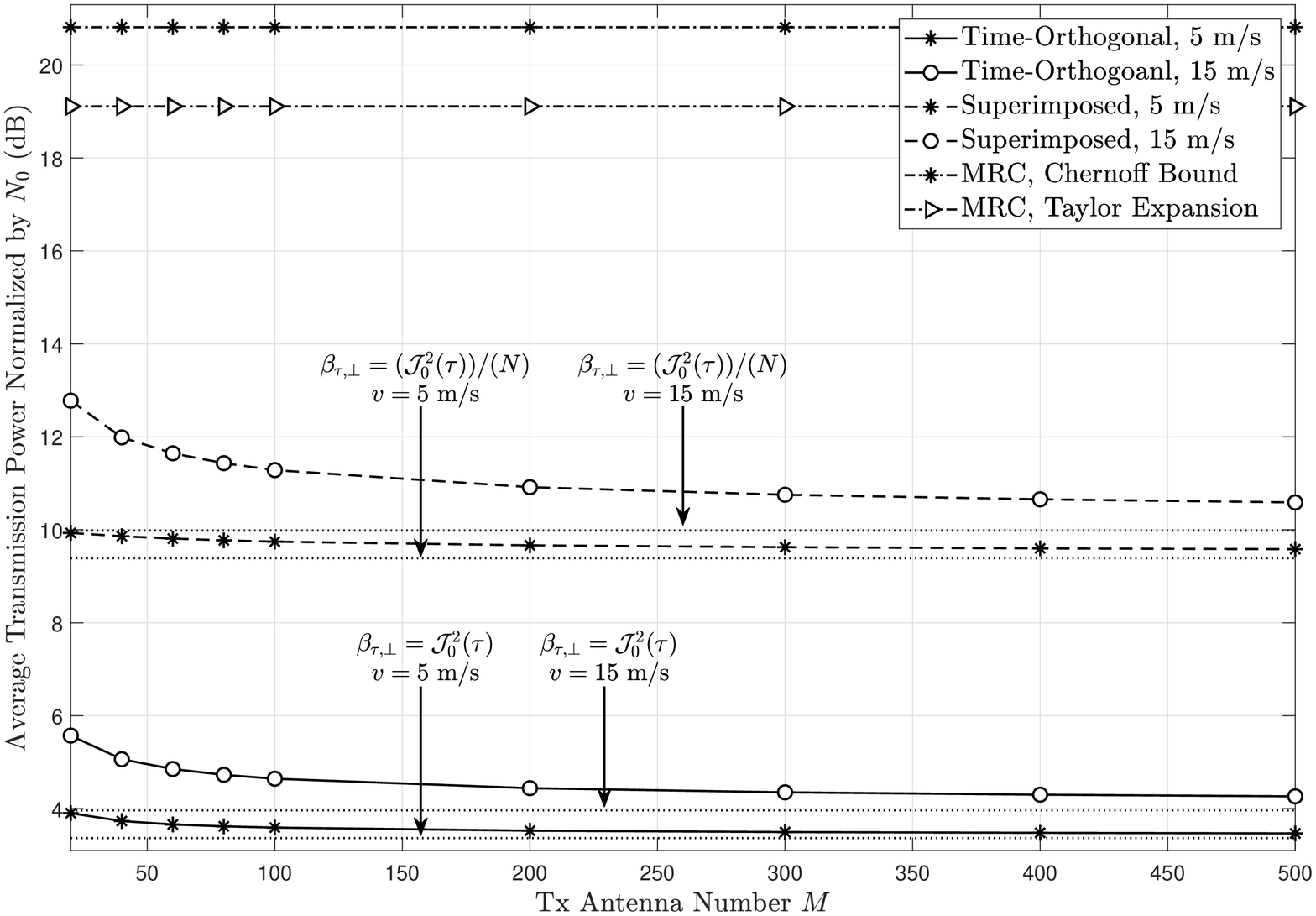}
\caption{The average transmission power as the Tx antenna number $M$ increases from $20$ to $500$ for different values of user velocity when the $N=4$ and $\mathscr{P}_\textsc{out}=0.2\times10^{-5}$.}
\vspace{-0em}
\label{fig6}
\end{minipage}
\begin{minipage}[t]{0.48\textwidth}
\captionsetup{width=0.96\textwidth}
\centering
\includegraphics[scale=0.29]{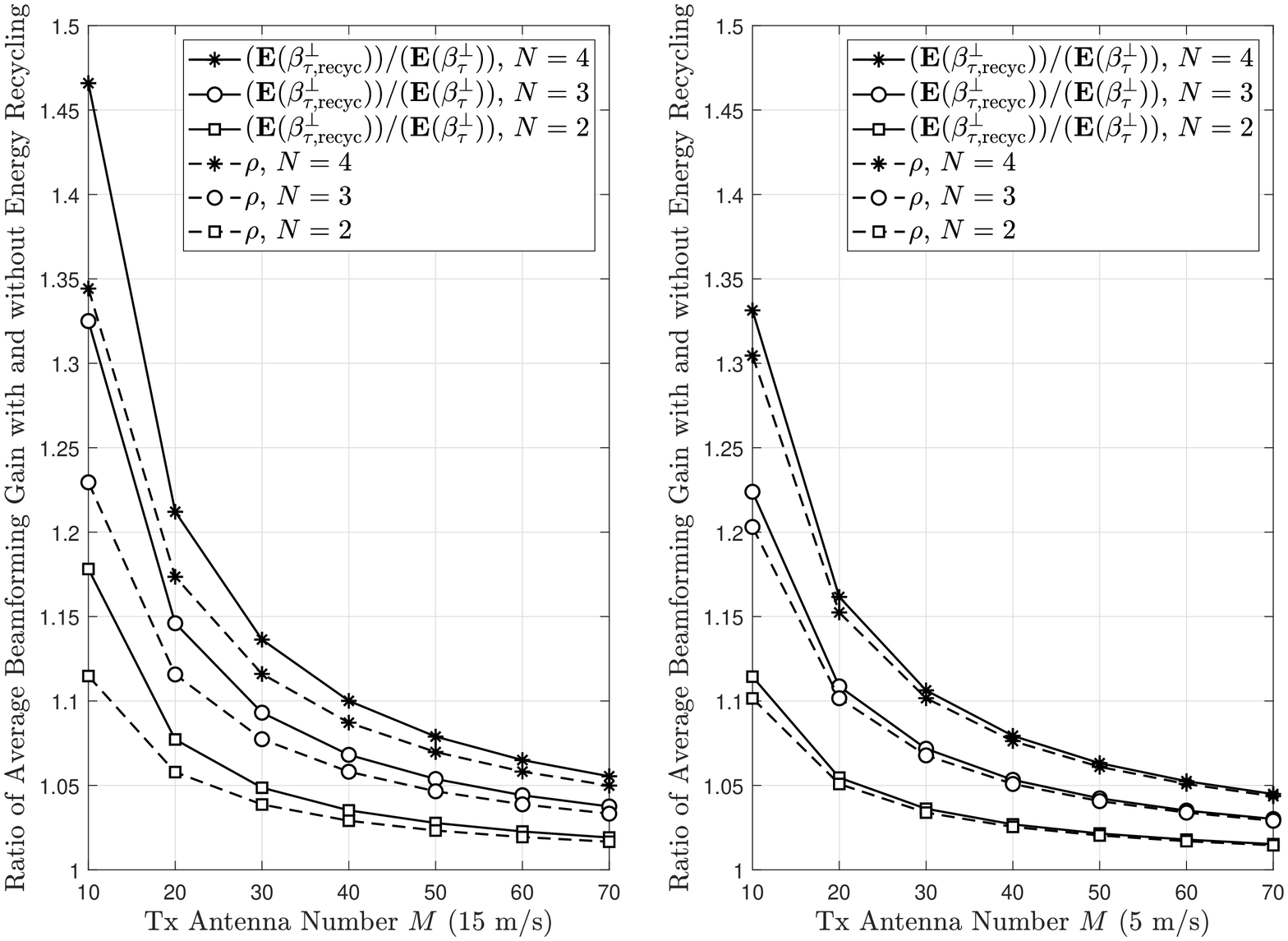}
\caption{The ratio of average beamforming gain brought by energy recycling as the Tx antenna number $M$ increases from $10$ to $70$ for different values of user velocity when $\mathscr{P}_\textsc{out}=0.2\times10^{-5}$.}
\vspace{-0em}
\label{fig7}
\end{minipage}
\end{figure}



The advantage of exploiting the Tx spatial diversity is demonstrated in Fig. \ref{fig6}. 
It can be observed that adopting the MF beamforming can bring significant improvement on the average transmission power. For the superimposed beamforming, there is more than $6$ dB gain on the average transmission power compared to using Taylor expansion for the MRC. While for the time-orthogonal beamforming, there is more than $13$ dB gain. This is reasonable, since the MRC can only exploit the Rx spatial diversity, which could be limited in many cases. While when adopting the MF beamforming, the massive Tx spatial diversity can help to improve the iSNR. 
It can also be observed that as $M$ increases, the average transmission power of the superimposed and time-orthogonal beamforming gradually approaches the case where $\beta_{\tau}^{\perp}=\mathcal{J}_0^2(\tau)$ and $\beta_{\tau}^{\perp}=(\mathcal{J}_0^2(\tau))/(N)$, respectively. This coincides with the results in Fig. \ref{fig4}. When the user velocity is $5$ m/s, there is only around $0.3$ dB gain for the average transmission power as $M$ increase from $20$ to $500$ for both superimposed and time-orthogonal beamforming. However, when the user velocity is $15$ m/s, there is more than $2$ dB gain as $M$ increases from $20$ to $500$. This is because when the user velocity is $5$ m/s, $\mathcal{J}_0$ is close to $1$. In this case, the efficiency to exploit the Tx spatial diversity is high and $\beta_{\tau}^\perp$ quickly approaches its convergence. While when the user velocity is $15$ m/s, the efficiency to exploit the spatial diversity is low, and it requires more Tx-antennas for $\beta_{\tau}^\perp$ to reach its convergence. This indicates that when the user velocity high, it requires higher diversity order of the Tx-antennas to improve the beamforming gain. 

The improvement brought by the energy recycling is demonstrated in Fig. \ref{fig7}. Specifically, the value of $(\mathbb{E}(\beta_{\tau,\text{recyc}}^\perp))/(\mathbb{E}(\beta_{\tau}^\perp))$ is demonstrated based on the numerical results in comparison with the measurement using aSNR (i.e., \eqref{ym047}).
It can be observed that, $(\mathbb{E}(\beta_{\tau,\text{recyc}}^\perp))/(\mathbb{E}(\beta_{\tau}^\perp))$ is only slightly higher than $\rho$ when the user velocity is $5$ m/s. But when the user velocity is $15$ m/s, the difference between $(\mathbb{E}(\beta_{\tau,\text{recyc}}^\perp))/(\mathbb{E}(\beta_{\tau}^\perp))$ and $\rho$ is significantly larger. This is reasonable, since as the user velocity approaches zero, the aged CSIT reduces to the perfect CSIT. In this case, $(\mathbb{E}(\beta_{\tau,\text{recyc}}^\perp))/(\mathbb{E}(\beta_{\tau}^\perp))$ is the same as $\rho$. Moreover, it can be observed that the energy recycling can bring around $10\%$ gain for $\mathbb{E}(\beta_{\tau}^\perp)$ when $M=40$ for both superimposed and time-orthogonal beamforming. This confirms our expectation that the energy harvesting can bring reasonable improvement for not-too-large massive-MIMO.

\begin{figure}[t]
\centering
\begin{minipage}[t]{0.48\textwidth}
\captionsetup{width=0.96\textwidth}
\centering
\includegraphics[scale=0.29]{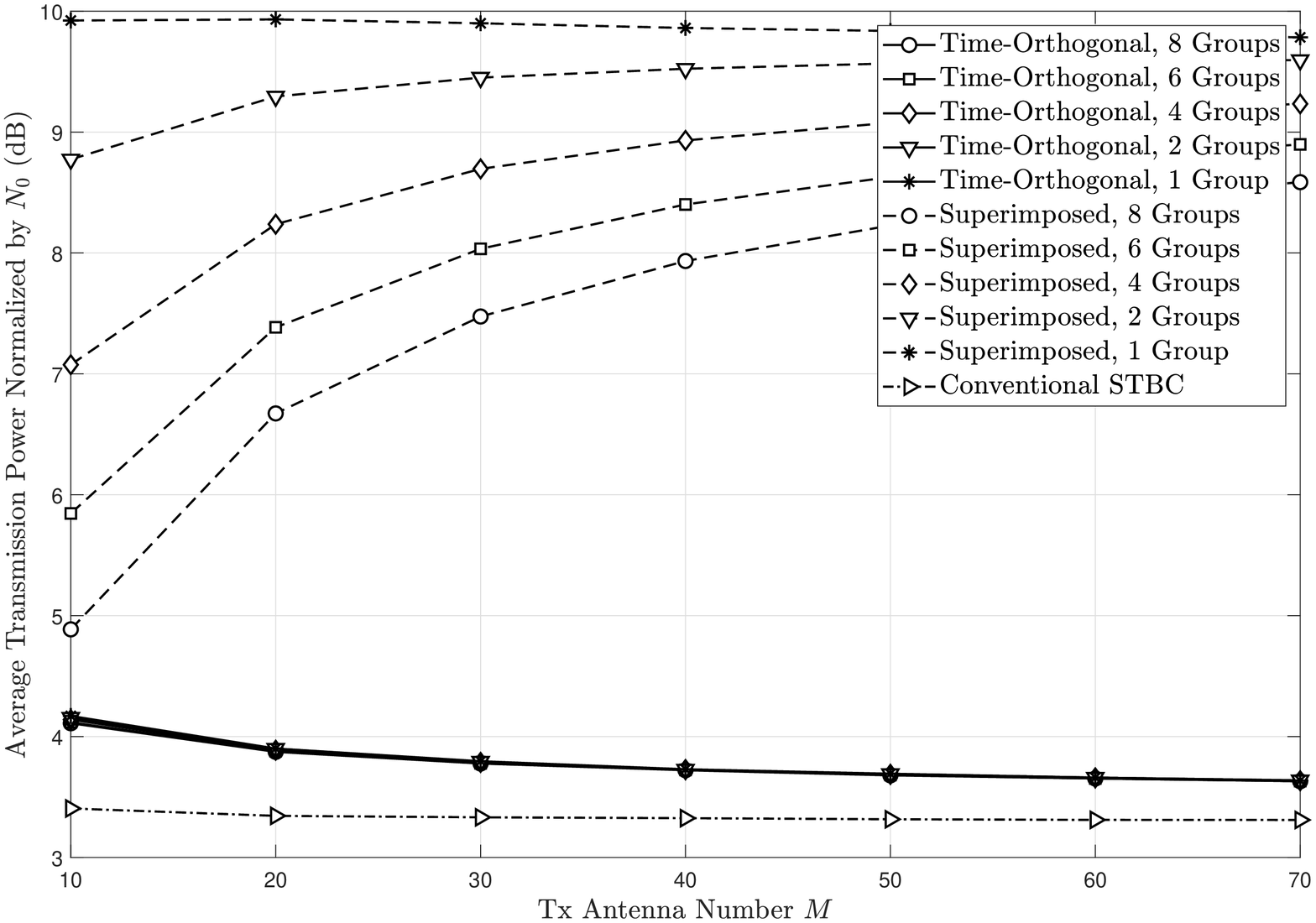}
\caption{The average transmission power as the Tx antenna number $M$ increases for both time-orthogonal and superimposed beamforming when $N=4$, $\mathscr{P}_\mathrm{out}=0.2\times10^{-5}$, and the user velocity is $5$ m/s.}
\vspace{-0em}
\label{fig8}
\end{minipage}
\begin{minipage}[t]{0.48\textwidth}
\captionsetup{width=0.96\textwidth}
\centering
\includegraphics[scale=0.29]{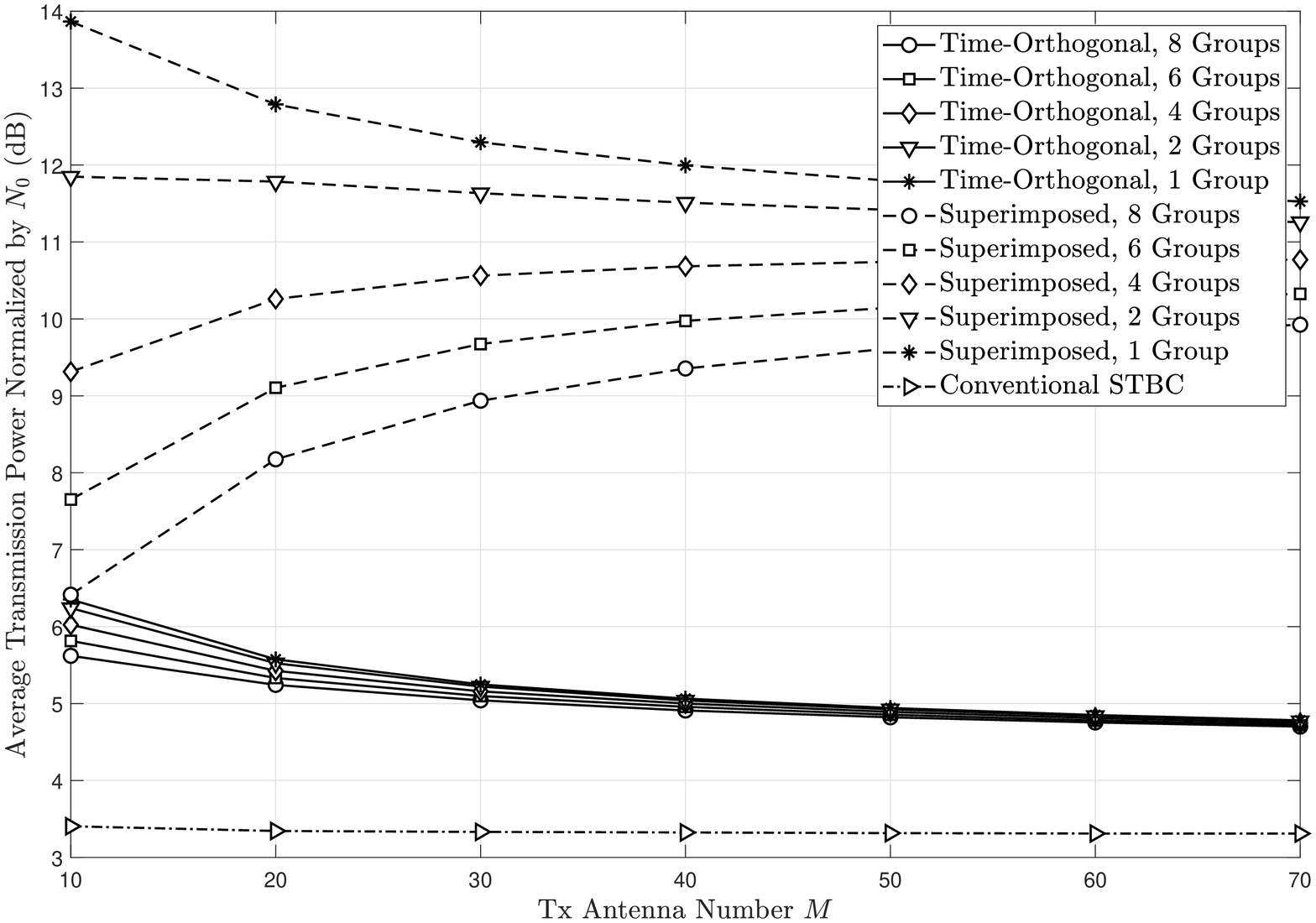}
\caption{The average transmission power as the Tx antenna number $M$ increases for both time-orthogonal and superimposed beamforming when $N=4$, $\mathscr{P}_\mathrm{out}=0.2\times10^{-5}$, and the user velocity is $15$ m/s.}
\vspace{-0em}
\label{fig9}
\end{minipage}
\end{figure}


\textit{Numerical Example 3:} 
The aim of this example is to demonstrate the improvement on average transmission power brought by the combinatorial approach of the MF beamforming and the G-STBC. 
Fig. \ref{fig8} shows the average transmission power of the combinatorial approach when the user velocity is $5$ m/s for both time-orthogonal and superimposed beamforming.
For the superimposed beamforming, the average transmission power can be significantly improved particularly when $M$ is small. Such improvement gradually vanishes as $M$ increases. This is because when $M$ increases, the channel within each antenna group gradually becomes orthogonal. In this case, the beamforming gain approaches the case in \textit{Corollary \ref{cor02}}, where $\beta_{\tau,\text{o}}^\perp$ is $N$ times of $\beta_{\tau,\text{s}}^\perp$. This indicates that the combinatorial can significantly improve the performance of the superimposed beamforming for not-too-large massive-MIMO. 
On the other hand, the combinatorial approach can hardly provide any improvement. But this is not surprising since when the user velocity is $5$ m/s, the CSIT uncertainty is limited.  
It can also be observed that the performance of the conventional STBC is only around $0.5$ dB better than the time-orthogonal beamforming. This can support the fact that the CSIT uncertainty is limited as well. 

Fig. \ref{fig9} shows the average transmission power when the user velocity is $15$ m/s. Similar to Fig. \ref{fig8}, the combinatorial approach can still significantly improve the performance of the superimposed beamforming when $M$ is not too large. The difference is that, when the grouping number is $8$, the combinatorial approach for the time-orthogonal beamforming can provide around $0.5$ dB gain when $M$ is less than $30$. It can also be observed that the performance improvement for the superimposed beamforming is higher than when the user velocity is $5$ m/s. For example, when $M=30$ and the group number is $8$, the improvement increases from $2.5$ dB ($5$ m/s) to $3.3$ dB ($15$ m/s). This indicates that the combinatorial approach can provide more improvement on the average transmission power when the user velocity is higher.

\section{Conclusion}\label{secVI}
In this paper, a novel pessimistic power adaptation approach has been proposed to enable URSST when MF beamforming is adopted for massive-MIMO. Specifically, a pessimistic bound of the beamforming gain was used to guarantee the outage requirement of every single transmission when the first-order Markov model is used to characterize the time-variant channel. Through our novel analysis, it has been revealed that the Chernoff lower bound has the advantage of tightness when MF beamforming is adopted. For the MF beamforming, it has also been revealed that the superimposed beamforming, which is popular in throughput-oriented systems, cannot maximize the iSNR for each individual Rx-antenna. Hence, the time-orthogonal MF beamforming has been adopted to improve the iSNR at the price of transmission latency. It has been proved that the time-orthogonal beamforming can significantly reduce the transmission power particularly when the Tx-antenna number is extremely high. 
The performance of the pessimistic power adaptation has been proved to be highly related to the time lag of the CSIT. In order to improve the beamforming gain when the lag is high, a novel combinatorial approach of the MF beamforming and the G-STBC has been proposed. It has been proved that the combinatorial approach can mitigate the impact of the CSIT uncertainty. Specifically for the superimposed beamforming, the combinatorial approach can significantly reduce the transmission power by decreasing the Tx-antenna number in each group for not-too-large massive-MIMO. Extensive computer evaluations in the i.i.d. Rayleigh-fading channel have been carried out to verify the conclusion above.

\appendices
\section{Proof of Theorem \ref{thm01}}\label{appdx1}
\begin{proof}
We first consider the case of the superimposed beamforming, where \eqref{ym027} can be expanded as:
\begin{equation}\label{apeq01}
f(t,\beta_{\tau}^\perp)=\exp(t\beta_{\tau}^\perp)\prod_{n=0}^{N-1}\mathbb{E}\left(\exp\left(-t\left\|\mathbf{h}_{n,\tau}^T\mathbf{w}\right\|^2\right)\right).
\end{equation}	
The term $\mathbf{h}_{n,\tau}^T\mathbf{w}$ conforms to $\mathcal{CN}\left(\mathbf{h}_{n,0}^T\mathbf{w},\sigma_\omega^2\right)$.
Then, we prove that for a random variable $\alpha\sim\mathcal{CN}(\mu_\alpha,\sigma_\alpha^2)$ ($\mu_\alpha\in\mathbb{C}$), the following result holds:
\begin{equation}\label{apeq02}
\mathbb{E}\left( \exp\left(-t|\alpha|^2\right) \right)=\frac{\exp\left(\frac{-|\mu_\alpha|^2t}{1+\sigma^2_\alpha t}\right)}{1+\sigma^2_\alpha t}.
\end{equation}
Since the real part and imaginary part of $\alpha$ have the same variance, $|\alpha|^2$ conforms to a non-central $\chi^2$ distribution with its PDF given by
\begin{IEEEeqnarray}{rl}\label{apeq03}
g(x)=\left\{
\begin{aligned}
&\sum_{i=0}^{\infty}\frac{\exp\left(\frac{-|\mu_\alpha|^2-x}{\sigma_\alpha^2}\right)\left(\frac{|\mu_\alpha|^2}{\sigma_\alpha^2}\right)^i x^i}{(i!)^2\sigma_\alpha^{2(i+1)}},&x>0\\
&0,&x\leq0
\end{aligned}
\right..
\end{IEEEeqnarray}
Hence, $\mathbb{E}\left( \exp\left(-t|\alpha|^2\right) \right)$ can be given by
\begin{equation}\label{apeq04}
\mathbb{E}\left( \exp\left(-t|\alpha|^2\right) \right)=\int_{0}^{\infty} e^{-tx}g(x)\mathrm{d}x.
\end{equation}
Since we have the following integration 
\begin{equation}\label{apeq05}
\int_{0}^{\infty}x^i\exp\left(-x\left(\frac{t\sigma_\alpha^2+1}{\sigma_\alpha^2}\right)\right)\mathrm{d}x=\left(\frac{\sigma_\alpha^2}{t\sigma_\alpha^2+1}\right)^{i+1}i!,
\end{equation} 
by substituting \eqref{apeq05} to \eqref{apeq04}, $\mathbb{E}\left( \exp\left(-t|\alpha|^2\right) \right)$ can be simplified as
\begin{equation}\label{apeq06}
\mathbb{E}\left( \exp\left(-t|\alpha|^2\right) \right)=\sum_{i=0}^{\infty}\frac{\exp\left(\frac{-|\mu_\alpha|^2}{\sigma_\alpha^2}\right)|\mu_\alpha|^{2i} }{i!\left(t\sigma_\alpha^2+1\right)^{i+1}\sigma_\alpha^{2i}}.
\end{equation}
Since we have the following convergence
\begin{equation}\label{apeq07}
\sum_{i=0}^{\infty}\frac{|\mu_\alpha|^{2i}}{i!\left(t\sigma_\alpha^2+1\right)^{i}\sigma_\alpha^{2i}}=\exp\left(\frac{|\mu_\alpha|^2}{(t\sigma_\alpha^2+1)\sigma_\alpha^2}\right),
\end{equation}
by substituting \eqref{apeq07} into \eqref{apeq06}, $\mathbb{E}\left( \exp\left(-t|\alpha|^2\right) \right)$ can be simplified as
\begin{equation}\label{apeq08}
\mathbb{E}\left( \exp\left(-t|\alpha|^2\right) \right)=\frac{\exp\left(\frac{-|\mu_\alpha|^2t}{1+\sigma_\alpha^2t}\right)}{1+\sigma_\alpha^2t}.
\end{equation}
Hence, by substituting $\mathbf{h}_{n,\tau}^T\mathbf{w}$ into $\alpha$, \eqref{apeq01} can be simplified as
\begin{equation}\label{apeq09}
f(t,\beta_{\tau}^\perp)=\frac{\exp\left(t\beta_{\tau}^\perp-\frac{(\mu_\text{s}(\tau)-N\sigma_\omega^2)t}{1+\sigma_\omega^2t}\right)}{(1+\sigma_\omega^2t)^{N}}.
\end{equation}
The mathematical form of \eqref{apeq09} also holds for the time-orthogonal beamforming. The only difference is that $\mu_\text{s}(\tau)$ is substituted for $\mu_\text{o}(\tau)$.  
\end{proof}

\section{Proof of Corollary \ref{cor02}}\label{appdx2}
\begin{proof}
We first consider the superimposed beamforming. Based on \eqref{ym021}, the limit of $\mu_\text{s}(\tau)$ after normalization is given by:
\begin{equation}\label{apeq10}
\lim\limits_{M\to\infty}\mu_\text{s}(\tau)=\frac{\mathcal{J}_0^2(\tau)}{N}\lim\limits_{M\rightarrow\infty}\frac{\left\|\mathbf{H}_{0}\mathbf{w}\right\|^2}{M}.
\end{equation}	
The term $\left\|\mathbf{H}_{0}\mathbf{w}\right\|^2$ can be expanded as
\begin{equation}\label{apeq11}
\left\|\mathbf{H}_0\mathbf{w}\right\|^2=\frac{\sum_{n_1=0}^{N-1}\big|\sum_{n_2=0}^{N-1}\mathbf{h}_{0,n_1}^T\mathbf{h}_{0,n_2}^*\big|^2}{\sum_{n_1=0}^{N-1}\sum_{n_2=0}^{N-1}\mathbf{h}_{0,n_1}^T\mathbf{h}_{0,n_2}^*}.
\end{equation}	
Since the channel is i.i.d Rayleigh fading, we have $\lim\limits_{M\rightarrow\infty}\mathbf{h}_{0,n_1}^T\mathbf{h}_{0,n_1}^*=M$ and $\lim\limits_{M\rightarrow\infty}\mathbf{h}_{0,n_1}^T\mathbf{h}_{0,n_2}^*=0$ ($n_1\neq n_2$).	
Then, it is trivial to prove
\begin{equation}\label{apeq12}
\lim\limits_{M\rightarrow\infty}\frac{\left\|\mathbf{H}_0\mathbf{w}\right\|^2}{M}=1.
\end{equation}

For the time-orthogonal beamforming, based on \eqref{ym023}, the limit of $\mu_\text{o}(\tau)$ is given by
\begin{equation}\label{apeq13}
\lim\limits_{M\to\infty}\mu_\text{o}(\tau)=\frac{\mathcal{J}_0^2(\tau)}{N}\lim\limits_{M\rightarrow\infty}\frac{\left\|\mathbf{H}_{0}\right\|^2}{M}.
\end{equation}
Since $\left\|\mathbf{H}_{0}\right\|^2=\sum_{n=0}^{N-1}\mathbf{h}_{0,n}^T\mathbf{h}_{0,n}^*$, it is trivial to prove	
\begin{equation}\label{apeq14}
\lim\limits_{M\rightarrow\infty}\frac{\left\|\mathbf{H}_0\right\|^2}{M}=N.
\end{equation}
By substituting \eqref{apeq12} and \eqref{apeq14} into \eqref{apeq10} and \eqref{apeq13}, respectively, \textit{Corollary \ref{cor02}} is proved.
\end{proof}

\section{Proof of Equation \eqref{eqn45}}\label{appdx3}
\begin{proof}
\eqref{ym047} can be expanded as
\begin{equation}\label{apeq15}
\rho\triangleq\frac{\sum_{n_1=0}^{N-1}\sum_{n_2=0}^{N-1}\mathbb{E}\left(\left\|\mathbf{h}_{n_1,\tau}^T\mathbf{w}_{n_2}\right\|^2 \right)}{\sum_{n=0}^{N-1}\mathbb{E}\left( \left\|\mathbf{h}_{n,\tau}^T\mathbf{w}_{n}\right\|^2  \right)}.
\end{equation}	
$\left\|\mathbf{h}_{n_1,\tau}^T\mathbf{w}_{n_2}\right\|^2$ conforms to a non-central $\chi^2$ distribution with its mean given by
\begin{equation}\label{apeq16}
\mathbb{E}\left(\left\|\mathbf{h}_{n_1,\tau}^T\mathbf{w}_{n_2}\right\|^2\right)=\mathcal{J}_0^2(\tau)\left\|\mathbf{h}_{n_1,0}^T\mathbf{w}_{n_2}\right\|^2+\sigma_\omega^2.
\end{equation}
To calculate the aSNR ratio, the ergodicity of the channel needs to be considered as well. Since the channel is i.i.d. Rayleigh fading, when $n_1=n_2$, we have
\begin{equation}\label{apeq17}
\mathbb{E}\left(\left\|\mathbf{h}_{n_1,0}^T\mathbf{w}_{n_2}\right\|^2\right)=M.
\end{equation}
When $n_1\neq n_2$, $\mathbf{h}_{n_1,0}$ is independent of $\mathbf{w}_{n_2}$:
\begin{IEEEeqnarray}{rl}
\mathbb{E}\left(\left\|\mathbf{h}_{n_1,0}^T\mathbf{w}_{n_2}\right\|^2\right)&=\sum_{m=0}^{M-1}\mathbb{E}\left(\left|h_{n_1,m,0}\right|^2\right)\mathbb{E}\left(\left|w_{n_2,m}\right|^2\right),~~~\label{apeq18}\\
&=\sum_{m=0}^{M-1}\mathbb{E}\left(\left|w_{n_2,m}\right|^2\right)=\left\|\mathbf{w}_{n_2}\right\|^2=1.\label{apeq19}
\end{IEEEeqnarray}
By substituting \eqref{apeq16}, \eqref{apeq17} and \eqref{apeq19} into \eqref{apeq15}, \eqref{eqn45} is proved.
\end{proof}

\ifCLASSOPTIONcaptionsoff
\newpage
\fi

\bibliographystyle{IEEEtran}
\bibliography{Bib_URLLC,Bib_Else}	

\end{document}